%% file: ugr_gri_paper.tex
\documentclass[useAMS,usenatbib]{mn2e}
\usepackage{graphicx}
\usepackage[english]{babel}
\usepackage{txfonts}
\usepackage{multirow}
\usepackage{lscape}
\usepackage{longtable}
\usepackage{natbib}
\usepackage{color}
\usepackage{float}

   \title{Investigating Emission Line Galaxy Surveys with the Sloan Digital Sky Survey Infrastructure}
   
   \author[Johan Comparat et al.]
   {Johan Comparat,$^1$
Jean-Paul Kneib,$^1$  Stephanie Escoffier,$^2$  Julien Zoubian,$^1$  Anne Ealet,$^2$  
\newauthor
Fabrice Lamareille,$^{3,4}$ N. Mostek,$^5$  Oliver Steele,$^6$ Eric Aubourg,$^{13}$  Stephen Bailey,$^5$   
\newauthor  Adam S. Bolton,$^{12}$ Joel Brownstein,$^8$ Kyle Dawson,$^5$ Jian Ge,$^{14}$ Olivier Ilbert,$^1$ 
\newauthor  Alexie Leauthaud,$^7$  Claudia Maraston,$^6$ Will Percival,$^6$ Nicholas P. Ross,$^5$ Carlo Schimd,$^1$  
\newauthor David J. Schlegel,$^5$ Donald P. Schneider,$^{9,10}$ Daniel Thomas,$^6$ Jeremy L. Tinker,$^{11}$ 
\newauthor Benjamin A. Weaver $^{11}$\\ 
$^1$Aix Marseille Universit\'e, CNRS, LAM (Laboratoire d'Astrophysique de Marseille) UMR 7326, 13388, Marseille, France\\
$^2$Centre de Physique des Particules de Marseille, Universit\'e d'Aix-Marseille, CNRS/IN2P3, Marseille, France\\
$^3$ Universit\'e de Toulouse; UPS-OMP; IRAP; Toulouse, France\\
$^4$ CNRS; IRAP; 14, avenue Edouard Belin, F-31400 Toulouse, France\\
$^5$ Lawrence Berkeley National Laboratory, One Cyclotron Road, Berkeley, CA 94720\\
$^6$ Institute of Cosmology and Gravitation (ICG), Dennis Sciama Building, Burnaby Road, Univ. of Portsmouth, Portsmouth, PO1 3FX, UK\\
$^7$ Institute for the Physics and Mathematics of the universe (IPMU), The University of Tokyo, Chiba 277-8582, Japan\\
$^8$ Department of Physics and Astronomy, University of Utah, Salt Lake City, UT 84112, USA\\
$^9$ Department of Astronomy and Astrophysics, The Pennsylvania State University,
  University Park, PA 16802\\
$^{10}$ Institute for Gravitation and the Cosmos, The Pennsylvania State University,
  University Park, PA 16802 \\
$^{11}$ Center for Cosmology and Particle Physics, New York University, New York, NY 10003 USA \\
$^{12}$Department of Physics and Astronomy, University of Utah, 115 South 1400 East, Salt Lake City, UT 84112, USA \\
$^{13}$APC, Univ. Paris Diderot, CNRS/IN2P3, CEA/Irfu, Obs de Paris, Sorbonne Paris Cit\'e, France \\
$^{14}$Department of Astronomy, University of Florida, USA.
  }

\begin{document}

\date{Accepted October 2nd 2012 by MNRAS. Received in original form July 17th 2012.}

\pagerange{\pageref{firstpage}--\pageref{lastpage}} \pubyear{2012}

\maketitle
\label{firstpage}

%----------------------abstract----------------------------%   
  \begin{abstract}
  The Baryon Acoustic Oscillation (BAO) feature in the power spectrum of galaxies provides a standard ruler to probe the accelerated expansion of the Universe.  The current surveys covering a comoving volume sufficient to unveil the BAO scale are limited to redshift $z \lesssim 0.7$. In this paper, we study several galaxy selection schemes aiming at building an emission-line-galaxy (ELG) sample in the redshift range $0.6<z<1.7$, that would be suitable for future BAO studies using the Baryonic Oscillation Spectroscopic Survey (BOSS) spectrograph on the Sloan Digital Sky Survey (SDSS) telescope. We explore two different colour selections using both the SDSS and the Canada France Hawai Telescope Legacy Survey (CFHT-LS) photometry in the \emph{u, g, r}, and \emph{i} bands and evaluate their performance selecting luminous ELG. From about 2,000 ELG, we identified a selection scheme that has a 75 percent redshift measurement efficiency. This result confirms the feasibility of massive ELG surveys using the BOSS spectrograph on the SDSS telescope for a BAO detection at redshift $z\sim1$, in particular the proposed \emph{eBOSS} experiment, which plans to use the SDSS telescope to combine the use of the BAO ruler with redshift space distortions using emission line galaxies and quasars in the redshift $0.6<z<2.2$.
   \end{abstract}
   
      
%--------------------------------------------------------------------------------------------------------------
%--------------------------------------------------------------------------------------------------------------
%------------------------------			INTRODUCTION       -----------------------------------
%--------------------------------------------------------------------------------------------------------------
%--------------------------------------------------------------------------------------------------------------
\section{Introduction}
\label{section:introduction}
%------------------ Table 1 BAO requirements ------------------------------%
\begin{table*}
	\caption{Requirements for measuring at 1 percent the BAO signal. 
$\bar{n}$ is the required density to overcome shot noise. The observed density [galaxy deg$^{-2}$] is the projection of $\bar{n}$ on the sky.
The Sample variance area required [deg$^2$] corresponds to the area necessary to obtain an effective volume of $1\;  \mathrm{Gpc}^3 \; h^{-3}$ at $k_1\simeq0.063\;h \; \mathrm{Mpc}^{-1}$ and $k_2\simeq0.12\;h \; \mathrm{Mpc}^{-1}$ given the value $\bar{n}$.
$N_\mathrm{req}$ is the number of thousands of redshifts required to detect BAO at $k_1$ and $k_2$: it is the product of the required observed density multiplied by the area required. `req.' stands for required.
%$N_\mathrm{req}=\mathrm{density} \; \mathrm{area}$ is the number of redshift in the redshift range necessary to sample the BAO in this redshift range at $k_1$ and $k_2$ (in thousands).
}
	\label{BAO_req}
	\centering
	\begin{tabular}{c c c c c c c c}
	\hline \hline	
		redshift	& \multicolumn{2}{c}{Shot Noise req.}				& \multicolumn{2}{c}{observed density req.} 	&Sample variance  & \multicolumn{2}{c}{$N_\mathrm{req}$} \\
		range	& $\bar{n}(k_{1})$ & $\bar{n}(k_{2})$ 				&\multicolumn{2}{c}{[deg$^{-2}$]}		&area req. [deg$^{2}$] & \multicolumn{2}{c}{[$10^{3}$ redshifts]} \\%	&  \\
				& \multicolumn{2}{c}{$10^{-4}h^3\mathrm{Mpc}^{-3}$}	& 	for $k_{1}$ & for $k_{2}$			&  	&  for $k_{1}$ & for $k_{2}$ \\
		\hline	
		$[0.3,0.6]$&1.0	&2.1	& 33  & 71  & 6188 		& 204  & 440 \\
		$[0.6,0.9]$&1.1	&2.5	& 75  & 162& 2585 		& 194  & 419 \\
		$[0.9,1.2]$&1.3	&2.9	& 121& 261& 1615 		& 195  & 421 \\
		$[1.2,1.5]$&1.5	&3.2	& 164& 354& 1227 		& 201  & 435 \\
		$[1.5,1.8]$&1.7	&3.6	& 273& 589& 1041 		& 284  & 613 \\\hline
	\end{tabular}
\end{table*}

With the discovery of the acceleration of the expansion of the universe \citep{1998AJ....116.1009R,1999ApJ...517..565P}, possibly driven by a new form of energy with sufficient negative pressure, recent results have concluded that $\sim96$ percent of the energy density of the universe is in a form not conceived by the Standard Model of particle physics and not interacting with the photons, hence dubbed ``dark''. Lying at the heart of this discovery is the distance-redshift relation mapped by the type Ia supernovae (SnIa) combined with the temperature power spectrum of the cosmic microwave background fluctuations. Since the first detections, there has been a huge increment of data up to redshift $z\sim 1$ (\citealt{1998AJ....116.1009R},\citealt{1999ApJ...517..565P},\citealt{2006A&A...447...31A},\citealt{2007ApJ...666..694W}, \citealt{2004ApJ...607..665R}, \citealt{2007ApJ...659...98R}, \citealt{2009AJ....138.1271D,Riess2011ApJ...730..119R})
The current precision and accuracy required to obtain deeper insight on the cosmological model using SnIa is limited by the systematic errors of this probe; therefore a joint statistical analysis with other probes is mandatory to assess a firm picture of the cosmological model.
%the utility of these supernovae datasets is now likely limited by systematic errors which could bias cosmological fits. It is highly desirable to cross-check the cosmological conclusions with other probes, as well as measure this distance-redshift relation to higher precision and larger redshift. 
%An increased precision of the distance scale must come through more robust, statistically-limited measurement techniques.
%One of the new attractive way to probe the distance-redshift relation is the measurement of the BAO scale imprinted in the large-scale structures probed by galaxies ({\it e.g.} \citealt{1998ApJ...504L..57E},\citealt{2003astro.ph..1623E},\citealt{2003ApJ...598..720S}, \citealt{2003PhRvD..68h3504L}). \citealt{2005MNRAS.362..505C}

Corresponding to the size of the well-established sound horizon in the primeval baryon-photon plasma before photon decoupling \citep{1970ApJ...162..815P}, the BAO scale provides a standard ruler allowing for geometric probes of the global metric of the universe. In the late-time universe it manifests itself in an excess of galaxies with respect to an unclustered (Poisson) distribution at the comoving scale $r \sim100 h^{-1} \mathrm{Mpc}$ --- corresponding to a fundamental wave mode $k\sim 0.063 h \mathrm{Mpc}^{-1}$. %({\it e.g.} \citealt{2005MNRAS.362..505C}, \citealt{2005ApJ...633..560E}). 
The value of this scale at higher redshift is accurately measured by the peaks in the CMB power spectrum ({\it e.g.} \citealt{2009ApJS..180..330K,Komatsu_2011}). Galaxy clustering and CMB observations therefore allow for a consistent comparison of the same physical scale at different epochs.
%The BAO scale corresponds to a feature imprinted in the distribution of photons and baryons by the propagation of sound waves in the relativistic plasma of the early universe ({\it e.g.} \citet{1970ApJ...162..815P, 1984ApJ...285L..45B, 1989ApJS...71....1H,1996ApJ...471..542H,1998ApJ...496..605E}). The BAO feature can be predicted very accurately by measurements of the CMB \citep{2009ApJS..180..330K,Komatsu_2011} and corresponds in the late-time galaxy clustering to a galaxy pair excess imprinted in the 2-point galaxy correlation function on typical scale of $r\sim100 {\rm Mpc}h^{-1}$ or $k\sim0.063 h {\rm Mpc^{-1}}$, which provides a cosmological standard ruler.

%BORING : The BAO feature was first observed in early 2005 both in the Sloan Digital Sky Survey (hereafter SDSS) Luminous Red Galaxy sample \citep{2005ApJ...633..560E}, and in the two degree Field Galaxy Redshift Survey (2dFGRS) data \citep{2005MNRAS.362..505C}; it is confirmed first in 2007 by using photometric Luminous Red Galaxies (LRGs) in 3,500 square degrees of SDSS \citep{2007MNRAS.378..852P}. Other concurrent clustering studies of the SDSS-II LRG sample were then carried out and confirmed the first measurement \citep{Percival_2010,Sanchez_2009,Chuang_2010,Kazin_2010,Reid_2010}. 
The first detection of the `local' BAO \citep{2005MNRAS.362..505C,2005ApJ...633..560E} 
%yield a value of \textcolor{red}{$r_\mathrm{BAO}=xx\pm yy$}, 
were based on samples at low redshift $z \leq 0.4$. Further analysis on a larger redshift range ($z>0.5$) and a wider area confirm the first result, reducing the errors by a factor of 2 \citep{Percival_2010,Blake_2011}.
%The first detection of the BAO \citep{2005MNRAS.362..505C,2005ApJ...633..560E} yield a value of \textcolor{red}{$r_\mathrm{BAO}=xx\pm yy$}, based on samples at low redshift $z \leq 0.4$. Further analysis on larger redshift range (z at XXX) and wider area ($XXXdeg^2$) confirm the first result reducing the errors by $xxx$ percent.
% 2dF $9.5\%$ on $\Omega_m$ with the power spectrum at $z\sim0.1$ 
% SDSS $8\%$ on $\Omega_m$ with the power spectrum at $z=0.35$ 
Measurements of the BAO feature have thus become an important motivation for large galaxy redshift surveys; the small amplitude of the baryon acoustic peak, and the large value of $r_\mathrm{BAO}$, require comoving volumes of order of $\sim 1 \mathrm{Gpc}^3 h^{-3}$ and at least $10^5$ galaxies to ensure a robust detection ({\it e.g.} \citealt{1997PhRvL..79.3806T,2003ApJ...594..665B}).

BAO studies using luminous red galaxies (LRG) are currently being pushed to $z=0.7$ by the Baryonic Oscillation Spectroscopic Survey (BOSS) experiment as part of the Sloan Digital Sky Survey III (SDSS-III) survey \citep{2011AJ....142...72E}. So far, with a third of the spectroscopic data, the BAO feature has been measured at $z=0.57$ with a $6.7\sigma$ significance \citep{BOSSDR9BAO2012arXiv1203.6594A}. The final data set, which will be completed by mid-2014, will have a mean galaxy density of about $150$ galaxies per square degree over 10,000 deg$^2$. Recently, the WiggleZ experiment has obtained a significant $\sim 4.9\sigma$ detection of the BAO peak at $z=0.6$, by combining information from three independent galaxy surveys: the SDSS, the 6-degree Field Galaxy Survey (6dFGS) and the WiggleZ Dark Energy Survey \citep{Blake_2011}. In contrast to SDSS, WiggleZ has mapped the less biased, more abundant emission line galaxies \citep{Drinkwater_2010}.

The next generation of cosmological spectroscopic surveys plans to map the high-redshift universe in the redshift range $0.6\leq z\leq2$ using the largest possible volume; see BigBOSS \citep{bigBOSS_2011}, PFS-SuMIRe\footnote{http://sumire.ipmu.jp/en/}, and EUCLID\footnote{http://sci.esa.int/euclid}.
% to probe the change between a dark matter dominated universe ( roughly $z>1$) and a dark energy dominated universe (roughly $z<1$) 
To achieve this goal, suitable tracers covering this redshift range are needed. Above $z \sim 0.6$ the number density of LRGs decreases while the bulk of galaxy population is composed of star forming galaxies \citep{Abraham_1996,Ilbert_06}; it is therefore compelling to build a large sample of such type of galaxies, which allows one to cover a large area and hence a large volume.
The main challenges for future BAO surveys is to efficiently select targets for which a secure redshift can be measured within a short exposure time. Contrary to continuum-based LRG survey, the observational strategy of next generation surveys such as BigBOSS, PFS-SuMIRe, and EUCLID is based on redshift measurements using emission lines, which are a common feature of star-forming galaxies. In this paper we focus on targeting strategies for selecting luminous ELGs at $0.6<z<1.7$ using optical photometry, and we test our strategies using the BOSS spectrograph on the SDSS telescope \citep{Gunn_2006}.

The plan of the paper is as follow. In section \ref{section:ELGs_BAO}, we derive the necessary ELG redshift distribution to detect the BAO feature. In section \ref{section:color_selection} we explain how the ELG selection criteria were designed using different photometric catalogs, based on the performances of the BOSS spectrograph. In section \ref{section:Measurements} we compare observed spectra issued from this selection with simulations and we discuss the efficiency of the proposed selection schemes. In section \ref{properties} we discuss the main physical properties of the ELGs. In section \ref{section:discussion}, we present the redshift distribution of the observed ELGs and how to improve the selection. In appendix \ref{tble_appendix} we display a representative set of the spectra observed.

Throughout this study we assume a flat $\Lambda$CDM cosmology characterized by 
$(\Omega_m, n_s, \sigma_8)=(0.27,0.96,0.81)$. Magnitudes are given in the AB system.
%$H_0=100 \; {\rm km}\;{\rm s}^{-1}\;{\rm Mpc}^{-1}$, $\Omega_m=0.27$, $\Omega_\Lambda=0.73$, $n_s=0.96$ $\sigma_8=0.81$. Magnitudes are given in the AB system.

%============== ============= ============= ========== 
%============= ============= ============= =========== 
%=============		ELGs and BAO			=============
%=============== ===================  ================
%=============== ===================  ================
\section{Baryon Acoustic Oscillations}
\label{section:ELGs_BAO}
%----------------- Figure 1: BAO needs ----------------------------
\begin{figure*}
\begin{center}
\includegraphics[width=180mm]{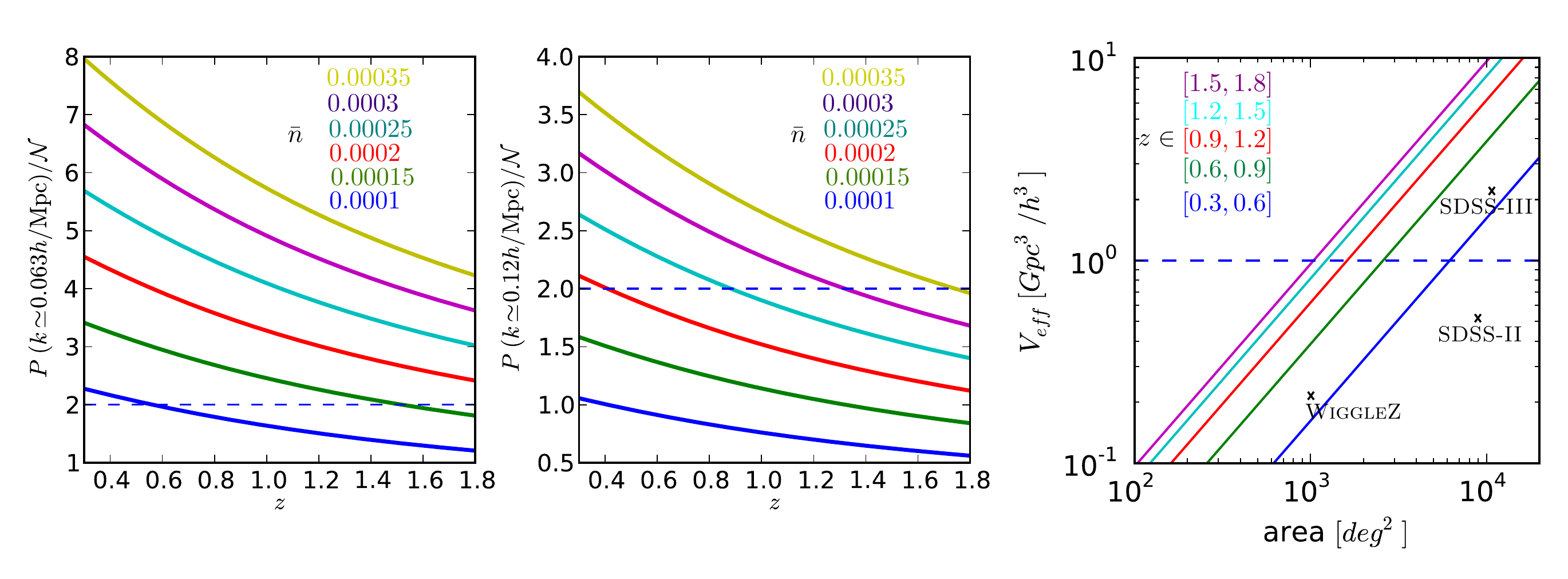}
\caption{$P/\mathcal{N}=\bar{n}P$, ratio of the fiducial power spectrum over the shot noise as a function of redshift at $k\simeq0.063$ {\bf(left)} and at $k\simeq0.12$ {\bf(center)} for $\bar{n}\in[1,3.5] \; 10^{-4}\; h^3\mathrm{Mpc}^{-3}$. {\bf (Right)} Effective volume in $\mathrm{Gpc}^3\;h^{-3}$ sampled as a function of sky coverage, for five different redshift bins from 0.3 to 1.8, the sampling density used for each bin is given in the column $\bar{n}(k_{2})$ of Table \ref{BAO_req}. To reach a comoving volume of $1\; \mathrm{Gpc}^3\;h^{-3}$ over the redshift range $[0.6,0.9]$, it is necessary to cover about $2500 \; \mathrm{deg}^2$.}
\label{ELG_bao_needs}
\end{center}
\end{figure*}

%\subsection{Requirements}
\subsection{Density and geometry requirements}
In order to constrain the distance-redshift relation at $z>0.6$ using the BAO, we need a galaxy sample that covers the volume of the universe observable at this redshift. In this section we derive the required mean number density of galaxy, $\bar{n}(z)$, and the area to be covered in order to observe the BAO feature at the one percent level.%Last sentence is redundant

The statistical errors in the measure of the power spectrum of galaxies $P(k,z)$, evaluated at redshift $z$ and at scale $k$, arise from sample variance and shot noise \citep{1986MNRAS.219..785K}. Denoting the latter as $\mathcal{N}(z)=1/\bar{n}(z)$, to measure a significant signal the minimal requirement is
\begin{equation}
\bar{n}(z)P(k, z) = \frac{P(k, z)}{\mathcal{N}(z)} \gtrsim 2.
\label{eqn:np_is_1}
\end{equation}
As the amplitude of the power spectrum decreases with redshift, the required density increases with redshift. {\it e.g.}, at $z=0.6$, we need a galaxy density of $\bar{n}=2.1 \times10^{-4}\; h^3\mathrm{Mpc}^{-3}$ ; at $z=1.5$, $\bar{n}=3.2 \times10^{-4}\; h^3\mathrm{Mpc}^{-3}$. The full trend in redshift bins is given in Table \ref{BAO_req} and in Figures \ref{ELG_bao_needs} a) and b) which show equation (\ref{eqn:np_is_1}) as a function of redshift for $k\simeq0.063 \;h  \; {\rm Mpc}^{-1}$ and $k\simeq0.12\;h \; {\rm Mpc}^{-1}$ (the location of the first and the second harmonics of the BAO peak in the linear power spectrum). 

In order to minimize the sample variance, we must sample the largest possible volume (a volume of 1 $\mathrm{Gpc}^3 \; h^{-3}$ roughly corresponds to a precision in the BAO scale measurement of 5 percent). To quantify this calculation, we use the effective volume sampled $V_{eff}$, defined as \citep{Tegmark_1997}
\begin{equation}
V_\mathrm{eff}(k)= 4 \pi \int dr \, r^2 \left[ \frac{\bar{n}(r) b^2(z) P(r,k)}{1+\bar{n}(r) b^2(z) P(r,k)} \right]^2 .
\label{eff_vol}
\end{equation}
In this calculation, we assume a linear bias according to the DEEP2 study by \citet{2008ApJ...672..153C} that varies according to the redshift as $b(z)=b_0 (1+z)$, with $b(z=0.8)=1.3$. The bias could be larger for the more luminous ELGs, that are thought to be the progenitors of massive red galaxies \citep{cooper2008}. We shall evaluate the bias of ELGs more precisely in a future paper. The corresponding area to be surveyed in order to reach $V_\mathrm{eff}\sim 1\mathrm{Gpc}^3 \; h^{-3}$ is shown in Table \ref{BAO_req}, setting redshift bins of width $\Delta z=0.3$ from $z=0.3$ to $z=1.8$. 
The Figure \ref{ELG_bao_needs} c) shows the behavior of $V_\mathrm{eff}$ as a function of the area for a given slice of redshift with $\bar{n}$ given in the third column of Table \ref{BAO_req}. For the redshift range $[0.6,0.9]$ the survey area must be $\gtrsim$2,500 $\mathrm{deg}^2$. For the redshift range $[0.9,1.2]$ the survey area must be $\gtrsim$1,600 $\mathrm{deg}^2$. The observation of  $[0.6,1.7]$ with a single galaxy selection thus needs 2,500 $\mathrm{deg}^2$ to sample the BAO at all redshifts.
%For a constant galaxy densities in bins of $\Delta z=0.3$ from $z=0.3$ to $z=1.8$, Table \ref{BAO_req} indicates the area to be surveyed to reach an effective volume of $1\mathrm{Gpc}^3 \; h^{-3}$. 

%\subsection{Reconstruction}
\subsubsection*{Reconstruction of the galaxy field}
%Nic Ross comment : Seems a bit much to dedicate a whole subsection to reconstruction, when it doesn't really relate to anything in the rest of the paper. I get the point that you want a higher space density to help here, but it's also not obvious (to me at least) whether an ELG, versus an LRG tracer helps you in mapping the density to displacement field. (I think this is a slightly tricky issue since bias, galaxy evolution, RSDs and e.g. Alcock-Paczynski all act in different ways - and signs! - with redshift...) 
To obtain a high precision on the measure of the BAO scale, it is necessary to correct the 2-point correlation function from the dominant non-linear effect of clustering. The bulk flows at a scale of $20\;h^{-1}\;{\rm Mpc}$ that form large scale structures smear the BAO peak: it is smoothed by the velocity of pairs  (At redshift 1 the rms displacement for biased tracers due to bulk flows is $8.5\;h^{-1}\; {\rm Mpc}$ in real space and $17\;h^{-1} \; {\rm Mpc}$ in redshift space) \citep{2007ApJ...664..675E,2007ApJ...664..660E}. 

Reconstruction consists in correcting this smoothing effect. The key quantity that allows reconstruction on a data sample is the smoothing scale used to reconstruct the velocity field and should be as close to $5\;h^{-1}\; {\rm Mpc}$ as possible in order to measure the bulk flows without being biased by other non-linear effects that occur on smaller scales. 

The reconstruction algorithm applied on the SDSS-II Data Release 7 \citep{Abazajian_2009} LRG sample sharpens the BAO feature and reduces the errors from 3.5 percent to 1.9 percent. This sample has a density of tracers of  $10^{-4}\; h^3\; {\rm Mpc}^{-3}$ and the optimum smoothing applied is $15\;h^{-1}\; {\rm Mpc}$ \citep{2012arXiv1202.0090P}. On the SDSS-III/BOSS data in our study (different patches cover 3,275 deg$^2$ on a total of 10,000 deg$^2$), reconstruction sharpens the BAO peak allowing a detection at high significance, but does not significantly improve the precision on the distance measure due to the gaps in the current survey (see \citealt{BOSSDR9BAO2012arXiv1203.6594A}). 

To allow an optimum reconstruction using a smoothing three times smaller ($5 \;h^{-1}\; {\rm Mpc}$) it is necessary to have a dense and contiguous galaxy survey : gaps in the survey footprint smaller than 1 Mpc and a sampling density higher than $3 \times 10^{-4}\; h^3 \; {\rm Mpc}^{-3}$. This setting should reduce the sample variance error on the acoustic scale by a factor four.% Therefore it should be possible to reach the percent level precision on the BAO determination % by combining 1 $\mathrm{Gpc}^3\;h^{-3}$ sampled at about $3 \times 10^{-4}\; h^3 \; {\rm Mpc}^{-3}$ with reconstruction, which will be our density goal for the target selections afterwards in the paper.

\subsection{Observational requirements}
A mean galaxy density of $3 \times 10^{-4}\; h^3 \; {\rm Mpc}^{-3}$ can be reached by a projected density of 162 galaxies $\mathrm{deg}^{-2}$ with $0.6<z<0.9$, 261 $\mathrm{deg}^{-2}$ with $0.9<z<1.2$, 354 with $1.2<z<1.5$, and 589 with $1.5<z<1.8$. Considering a simple case where a survey is divided in three depths, the shallow one covering 2,500 deg$^2$ should contain 419,000 galaxies ; the medium 421,000 galaxies over 1,600 deg$^{2}$ ; and the deep 435,000 galaxies over 1,200 deg$^{2}$. This represents a survey containing 1,350,000 measured redshifts in the redshift range $[0.6,1.5]$. The challenge is to build a selection function that enhances the observation of these projected densities. 

Given a ground-based large spectroscopic program that measures $1.5 \times10^6$ spectra (it corresponds to about 4 years of dark time operations on SDSS telescope dedicated to ELGs), the challenge is to define a selection criterion that samples galaxies to measure the BAO on the greatest redshift range possible. We define the selection efficiency as the ratio of the number of spectra in the desired redshift range and the number of measured spectra. The example in the previous paragraph needs a selection with an efficiency of $1.35/1.5\sim$ 90 percent. 
%Another example: a survey sampling the redshift range $[0.6,1.2]$ on 2500 deg$^{2}$ with a galaxy density of $2.9 \times 10^{-4}\; h^3 \; {\rm Mpc}^{-3}$, needs 1,057,500 redshifts and thus an efficiency around 70 percent. 
 
%\subsection{Previous galaxy targets selections}
\subsection{Previous galaxy targets selections}
To reach densities of tracers $\gtrsim10^{-4}\; h^3\; {\rm Mpc}^{-3}$ at $z>0.6$ with a high efficiency, a simple magnitude cut is not enough. Such a selection would be largely dominated by low-redshift galaxies. The use of colour selections is necessary to narrow the redshift range of the target selection for observations. 

SDSS-I/II galaxies are selected with visible colours in the red end of the colour distribution of galaxies, resulting in a sample of LRG and not ELGs \citep{2001AJ....122.2267E}. The projected density of LRG is $\sim120$ deg$^{-2}$ with a peak in the redshift distribution at $z\sim0.35$. With the SDSS-I/II LRG sample, the distance redshift relation was reconstructed at 2 percent at $z=0.35$. 

BOSS has currently completed about half of its observation plan. The tracers used by BOSS are, as SDSS-I/II LRG, selected in the red end of the color distribution of galaxies, they are called CMASS (it stands for `constant mass' galaxies) and the selection will be detailed in Padmanabhan et al. in prep. (2012). The current BAO detection using the data release 9 (a third of the observation plan) with the CMASS tracers at $z\sim 0.57$ has a $6.7\sigma$ significance (\citealt{BOSSDR9BAO2012arXiv1203.6594A}).

WiggleZ blue galaxies are selected using UV and visible colours: they have a density of 240 galaxies deg$^{-2}$ and a peak in the redshift distribution around $z=0.6$ \citep{Drinkwater_2010}. The WiggleZ experiment has obtained a $4.9\sigma$ detection of the BAO peak at $z=0.6$ \citep{Blake_2011}.

At their peak density, both of these BAO surveys reach a galaxy density of $3 \times 10^{-4}\; h^3\; {\rm Mpc}^{-3}$, which guarantees a significant detection of the BAO. 

Galaxy selections beyond $z=0.6$ were already performed by surveys such as the VIMOS-VLT Deep Survey\footnote{http://cesam.oamp.fr/vvdsproject/} (VVDS, see \citealt{2005Natur.437..519L}), DEEP2\footnote{http://deep.berkeley.edu/index.html} (see \citealt{Davis_2003}) or Vimos Public Extragalactic Redshift Survey\footnote{http://vipers.inaf.it/project.html} (VIPERS, see Guzzo et al. 2012, in preparation), but they are not tuned for a BAO analysis. The DEEP2 Survey selected galaxies using BRI photometry in the redshift range $0.75-1.4$ on a few square degrees with a redshift success of 75 percent using the Keck Observatory. It studied the evolution of properties of galaxies and the evolution of the clustering of galaxies compared to samples at low-redshift. In particular, insights in galaxy clustering to $z=1$ brings a strong knowledge about the bias of these galaxies \citep{2008ApJ...672..153C}. The VVDS wide survey observed 20 000 redshift on 4 deg$^2$ limited to $I_{AB}<22.5$ \citep{2008A&A...486..683G}; they studied the properties of the galaxy population to redshift $1.2$ and the small scale clustering around $z=1$. %The absence of a colour cut implies that most galaxies are below a redshift of $0.6$. 
The VIPERS survey maps the large scale distribution of 100 000 galaxies on 24 $\mathrm{deg}^2$ in the redshift range $0.5-1.2$ to study mainly clustering and redshift space distortions. Their colour selection, based on \emph{ugri} bands, is described in more detail in the section \ref{section:discussion}.

%============== ============= ============= ========== 
%============= ============= ============= =========== 
%=============		   colour SELECTION     =============
%=============== ===================  ================
%=============== ===================  ================
\section{Color Selections}
\label{section:color_selection}

Our aim is to explore different colour selections that focus on galaxies located in $0.6<z<1.7$ with strong emission lines, so that assigning redshifts to these galaxies is feasible within short exposure times (typically one hour of integration on the 2.5m SDSS telescope). The methodology used here has been first explored and experimented by \citet{Davis_2003}, \citealt{Adelberger_2004}, \citealt{Drinkwater_2010}. \citet{Adelberger_2004} derived different colour selections for faint galaxies (with $23<R<25.5$) at redshifts $1<z<3$ based on the Great Observatories Origins Deep Survey data (GOODS, see \citealt{2003ApJ...587...25D}). \citet{Drinkwater_2010} selected ELGs using UV photometry from the Medium Imaging Survey of the GAlaxy EVolution EXplorer (MIS-GALEX, see \citealt{2005ApJ...619L...1M}) data combined with SDSS, to obtain a final density of $238$ ELGs per square degree with $0.2< z <0.8$ over $\sim 800$ square degrees.

Our motivation is to probe much wider surveys than GOODS or GALEX (ultimately a few thousands square degrees) and to concentrate on intrinsically more luminous galaxies (typically with $g<23.5$) %\textcolor{red}{Quote B-band restframe luminosity?} 
with a redshift distribution extended to redshift 1.7.

The selection criteria studied in this work are designed for a ground-based survey and more specifically for the SDSS telescope, a 2.5m telescope located at Apache Point Observatory (New Mexico, USA), which has a {\it unique} wide field of view to carry out LSS studies \citep{Gunn_2006}. The current BOSS spectrographs cover a wavelength range of $3600-10200 \AA$. Its spectral resolution, defined by the wavelength divided by the resolution element, varies from $R\sim 1,600$ at $3,600\AA$ to $R\sim 3,000$ at $10,000\AA$ \citep{2011AJ....142...72E}. The highest redshift detectable with the $\left[\mathrm{O\textrm{\textsc{ii}}}\right]$ emission line doublet $(\lambda 3727,\lambda 3729)$ is thus $z_\mathrm{max}=1.7$.

To select ELGs in the redshift range $[0.6,1.7]$ we have explored two different selection schemes: first using \emph{u,g,r} photometry and secondly using \emph{g,r,i} photometry.% We then test these selections using real observations on the SDSS telescope. We first briefly describe the SDSS and the CFHT-LS surveys we will use to build the colour selections.

%\subsection{Photometric data properties: SDSS, CFHT-LS and COSMOS photo-zs}
\subsection{Photometric data properties: SDSS, CFHT-LS and COSMOS}% photo-zs}
%------------ Figure 2: CFHT SDSS mag errors  ----------------------%
\begin{figure}
\begin{center}
\includegraphics[width=88mm]{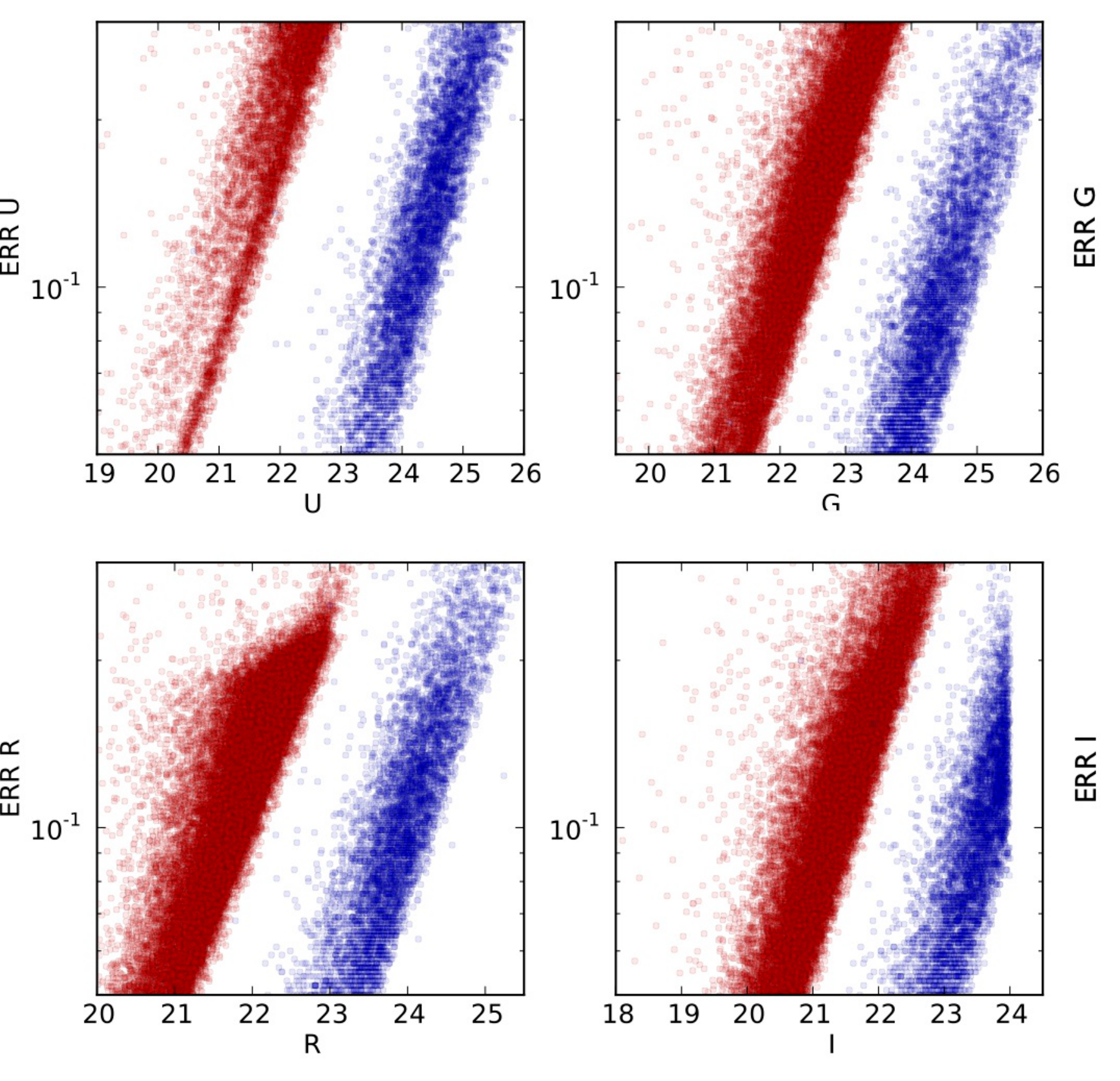}%figures/err_mag_sdss_cfht.pdf}
\caption{The four bands \emph{ugri} and their precision are illustrated; in red for SDSS photometry; in blue for CFHTLS photometry. The \emph{u} band quality is limiting the precision of the colour selection on SDSS photometry. Note that the photometric redshift CFHTLS catalog is cut at $i=24$, and the SDSS data is R-selected with $err_R\leq 0.2$.}
\label{mag_errors_SDSS_CFHT}
\end{center}
\end{figure}

%------------------------- Figure 3 : selection ugr, gri, --------------------------%
\begin{figure*}
\begin{center}
\includegraphics[width=150mm]{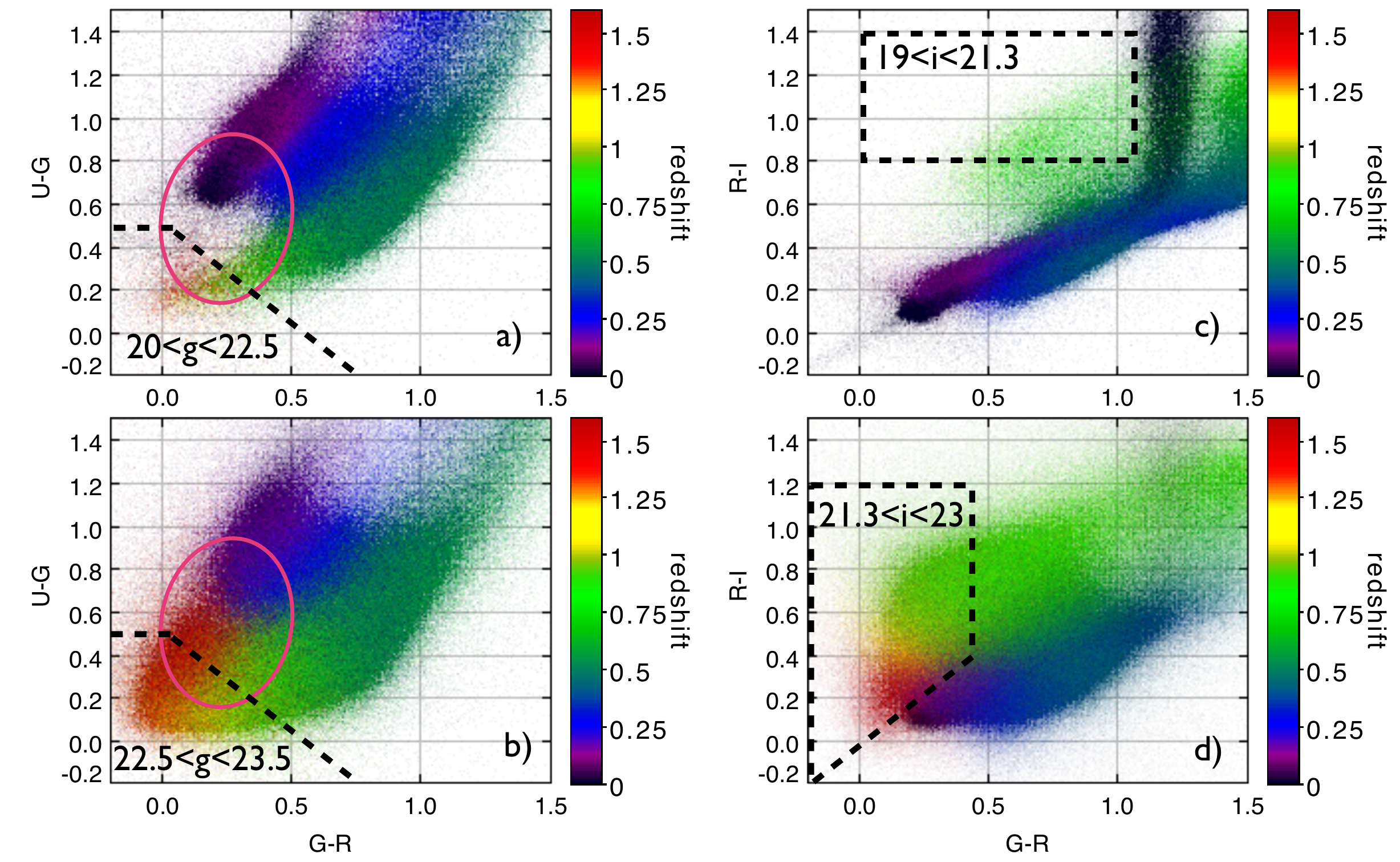}
\caption{Color-color diagrams tinted according to CFHT-LS photometric redshift. Colour selections are indicated with the dashed boxes. Quasars, when overlapping with the colour selection, are located in the pink ellipse. The stellar sequence appears in black. \textbf{a)} The bright \emph{ugr} colour selection: \emph{u-g} vs \emph{g-r} for $20<g<22.5$. In this area the overall density of targets is low.
\textbf{b)} The faint \emph{ugr} colour selection where the magnitude range is $22.5<g<23.5$. The density of available targets increases and the stellar sequence is diminished.
\textbf{c)} The bright \emph{gri} colour selection: \emph{r-i} vs \emph{g-r} diagram for $19<i<21.3$. The high-redshift targets are not as visible as in the \emph{ugr} plot. The target density is also small.
\textbf{d)} The faint \emph{gri} colour selection where the magnitude range is $21.3<i<23$. The target density increases and higher redshift targets appear in the blue end of the colour plot while the stellar locus fades.
Using SDSS photometry one would obtain similar plots to \textbf{a)} and \textbf{c)} in terms of target density.}
\label{selection_figure}
\end{center}
\end{figure*}

The photometric SDSS survey, delivered under the data release 8 (DR8, \citealt{SDSS_DR8}), covers 14,555 square degrees in the 5 photometric bands \emph{u, g, r, i, z}. It is the largest volume multi-color extragalactic photometric survey available today. The 3$\sigma$ magnitude depths are: $u=22.0$, $g=22.2$, $r=22.2$, $i=21.3$; see \citet{1996AJ....111.1748F} for the description of the filters and \citet{1998AJ....116.3040G} for the characteristics of the camera. The magnitudes we use are corrected from galactic extinction.

The Canada France Hawaii Telescope Legacy Survey\footnote{http://www.cfht.hawaii.edu/Science/CFHLS/} (hereafter CFHTLS) covers $\sim155$ deg$^2$ in the \emph{u,g,r,i,z} bands. The transmission curves of the filters differ slightly\footnote{http://cadcwww.dao.nrc.ca/megapipe/docs/filters.html} from SDSS. The  data and cataloging methods are described in the T0006 release document\footnote{http://terapix.iap.fr/cplt/T0006-doc.pdf}. The 3$\sigma$ magnitude depths are: $u=25.3$, $g=25.5$, $r=24.8$, $i=24.5$. The CFHT-LS photometry is ten times (in $r$ and $i$) to thirty times (in $u$) deeper than SDSS DR8, however the CFHTLS covers a much smaller field of view than SDSS DR8. The magnitudes we use are corrected from galactic extinction. The CFHT-LS photometric redshift catalogs are presented in \citet{Ilbert_06}, and \citet{Coupon_2009} ; the photometric redshift accuracy is estimated to be $\sigma_z < 0.04 (1+z)$ for $g\leq 22.5$. This photometric redshift catalog is cut at $i=24$, beyond which photometric redshifts are highly unreliable. Fig. \ref{mag_errors_SDSS_CFHT} displays the relative depth between SDSS and CFHT-LS wide surveys in the \emph{u,g,r,i}-bands.

COSMOS is a deep 2 deg$^2$ survey that has been observed at more than 30 different wavelengths \citep{2007ApJS..172....1S}. The COSMOS photometric catalog is described in \citet{Capak_2007} and the photometric redshifts in \citet{Ilbert_2009}. The COSMOS Mock Catalog, (hereafter CMC; see\footnote{http://lamwws.oamp.fr/cosmowiki/RealisticSpectroPhotCat}) is a simulated spectro-photometric catalog based on the COSMOS photometric catalog and its photometric redshift catalog. The magnitudes of an object in any filter can be computed using the photometric redshift best-fit spectral templates %to which noise is added \textcolor{red}{About the noise, Explain more or Remove} 
(\citealt{Jouvel_2009}, Zoubian et al. 2012, in preparation).

The limiting magnitudes of the CMC in the each band are the same as in the real COSMOS catalog (detection at $5\sigma$ in a 3" diameter aperture): $\emph{u}<26.4$, $\emph{g}<27$, $\emph{r}<26.8$, $\emph{i}<26.2$. For magnitudes in the range $14<m<26$ in the \emph{g,r,i} bands from the Subaru telescope and in the \emph{u} band from CFHTLS, the CMC contains about 280,000 galaxies in 2 deg$^2$ to COSMOS depth. 
The mock catalog also contains a simulated spectrum for each galaxy. These simulated spectra are generated with the templates used to fit COSMOS photometric redshifts. Emission lines are empirically added using Kennicutt calibration laws \citep{Kennicutt_1998,Ilbert_2009}, and have been calibrated using zCOSMOS \citep{Lilly_2009} as described in Zoubian et al. 2012, in preparation. The strength of $\left[\mathrm{O\textrm{\textsc{ii}}}\right]$ emission lines was confirmed using DEEP2 and VVDS DEEP luminosity functions \citep{LeFevre2005,zhu09}. Finally a host galaxy extinction law is applied to each spectrum. Predicted observed magnitudes take into account the presence of emission lines.

%\subsection{ugr and gri selections}
\subsection{Color selections}

Based on the COSMOS and CFHT-LS photometric redshifts, we explore two simple colour selection functions using the \emph{ugr} and \emph{gri} bands. Fig. \ref{selection_figure} shows the targets available in the \emph{ugr} and \emph{gri} colour planes. We construct a bright and a faint sample based on the photometric depths of SDSS and CFHT-LS. 

\subsubsection{{\it ugr} selection}
The \emph{ugr} colour selection is defined by $-1<u-r<0.5$ and $-1<g-r<1$ that selects galaxies at $z\geq 0.6$ and ensures that these galaxies are strongly star-forming ($u-r$ cut). The cut $-1<u-g<0.5$ removes all low-redshift galaxies ($z<0.3$). Finally the magnitude range is $20<g<22.5$ and $g<$23.5 for the bright and the faint samples, resp. Fig. \ref{selection_figure} a) and b).

\subsubsection{{\it gri} selection}
The bright \emph{gri} colour selection is defined by the range $19<i<21.3$. We select blue galaxies at $z\sim 0.8$ with $0.8<r-i<1.4$ and $-0.2<g-r<1.1$ (Fig. \ref{selection_figure} c). In the faint range $21.3<i<23$, we tilt the selection to select higher redshifts with $-0.4<g-r<0.4$, $-0.2<r-i<1.2$ and $g-r<r-i$ (Fig. \ref{selection_figure} d).

%------------------------- Table 2: all Samples --------------------------%
\begin{table*}
\caption{Properties of the colour-selected samples.
`b' stands for bright ($20<g<22.5$ for \emph{ugr} or $19<i<21.3$ for \emph{gri}) and `f' for faint ($22.5<g<23.5$ for \emph{ugr} or $21.3<i<23$ for \emph{gri}).
We indicate the mean and standard deviation of the redshift distribution for each set. 
For the CMC sample, $\left[\mathrm{O\textrm{\textsc{ii}}}\right]$ fluxes are expressed in units of $10^{-17} \mathrm{erg\,cm^{-2}\,s^{-1}}$ ; the median, the first and third quartiles of the flux distribution are indicated. The \emph{ugr} tends to select the highest $\left[\mathrm{O\textrm{\textsc{ii}}}\right]$ fluxes.}
	\label{mock_selections}
	\centering
	\begin{tabular}{c c c c c c c c c c c c c}
	\hline \hline
	\multicolumn{3}{c}{selection} &  $\# \mathrm{deg}^{-2}$ & $\bar{u}$& $\bar{g}$& $\bar{r}$ & $\bar{i}$& $\bar{z}$ & $\sigma_z $& $\bar{f_{\left[\mathrm{O\textrm{\textsc{ii}}}\right]}}$ &$Q^1_{f_{\left[\mathrm{O\textrm{\textsc{ii}}}\right]}}$  &$Q^3_{f_{\left[\mathrm{O\textrm{\textsc{ii}}}\right]}}$\\
	\hline
	\multirow{4}{*}{CMC} & \multirow{2}{*}{\emph{ugr}} 	& b &  130.0  &  21.98   &  21.87  &   21.69  &    -  &  1.25  &  0.53  &  61.74  &  46.47 &88.39\\
					& 					    	& f &  1450.8  &  23.27  &  23.18  &   22.98  &   -  &    1.19  &  0.38  &  16.60  &  13.06 & 22.26
\\
					& \multirow{2}{*}{\emph{gri}} 	& b &  257.2  &  -  	   &  22.69  &  21.87  &    20.93  &   0.80  &  0.21  &  13.85  &  8.65 & 22.21\\
					& 						& f &  2170.5  &  -   	   &  23.34  &  23.09  &    22.55  &   0.93  &  0.31  &  10.23  &  6.83 & 15.99
\\
	\hline
	\multirow{4}{*}{CFHT-W1} & \multirow{2}{*}{\emph{ugr}} & b & 193.3  &  21.95   &  21.8    &  21.7   &  -  &  1.28  &  0.38\\
					& 					       & f &1766.8  &  23.37    &  23.19    &  23.07  &  -   &  1.29  & 0.31\\
					& \multirow{2}{*}{\emph{gri}}    & b & 361.4  &  -  &  22.62   &  21.8  &  20.82  &   0.81  &  0.11\\
					& 					       & f & 3317.5 &  -  &  23.34    &  23.11  &  22.55  &    1.03  &  0.35\\
	\hline
	\multirow{4}{*}{CFHT-W3} & \multirow{2}{*}{\emph{ugr}} &b &232.2  &  21.89  &  21.76  &     21.69  &    -    &  1.27  &  0.37\\
					& 					       & f &1679.1  &  23.36  &    23.18  &     23.06    &  -  &  1.28  &  0.31\\
					& \multirow{2}{*}{\emph{gri}}    & b &391.6  &  -  & 22.62  &  21.78  & 20.8  &    0.82  &  0.1\\
					& 					       &f &3334.2  &  -  & 23.34  &    23.11  &   22.54  &   1.03  &  0.33\\
	\hline
	\multirow{2}{*}{SDSS} & \multirow{1}{*}{\emph{ugr}}	 & b &166.96  &  21.76  &     21.77  &     21.52  &     - \\
%					&						 & $g<23.5$ & 367.28  &  22.25  &  0.65  &  22.41  &  0.75  &  22.0  &  0.64  &  -  &  -  \\
					& \multirow{1}{*}{\emph{gri}} 	& b & 204.96  &  -  &  22.57  &  21.75  &   20.76 \\
%					&						 & $i<23.0$ &322.56  &  -  &  -  &  22.82  &  0.55  &  22.04  &  0.54  &  21.07  &  0.54 \\
	\hline 	
	\end{tabular}
\end{table*}

%\subsection{Properties of galaxy samples}
\subsection{Predicted properties of the selected samples}

\begin{figure*}
\begin{center}
\includegraphics[width=180mm]{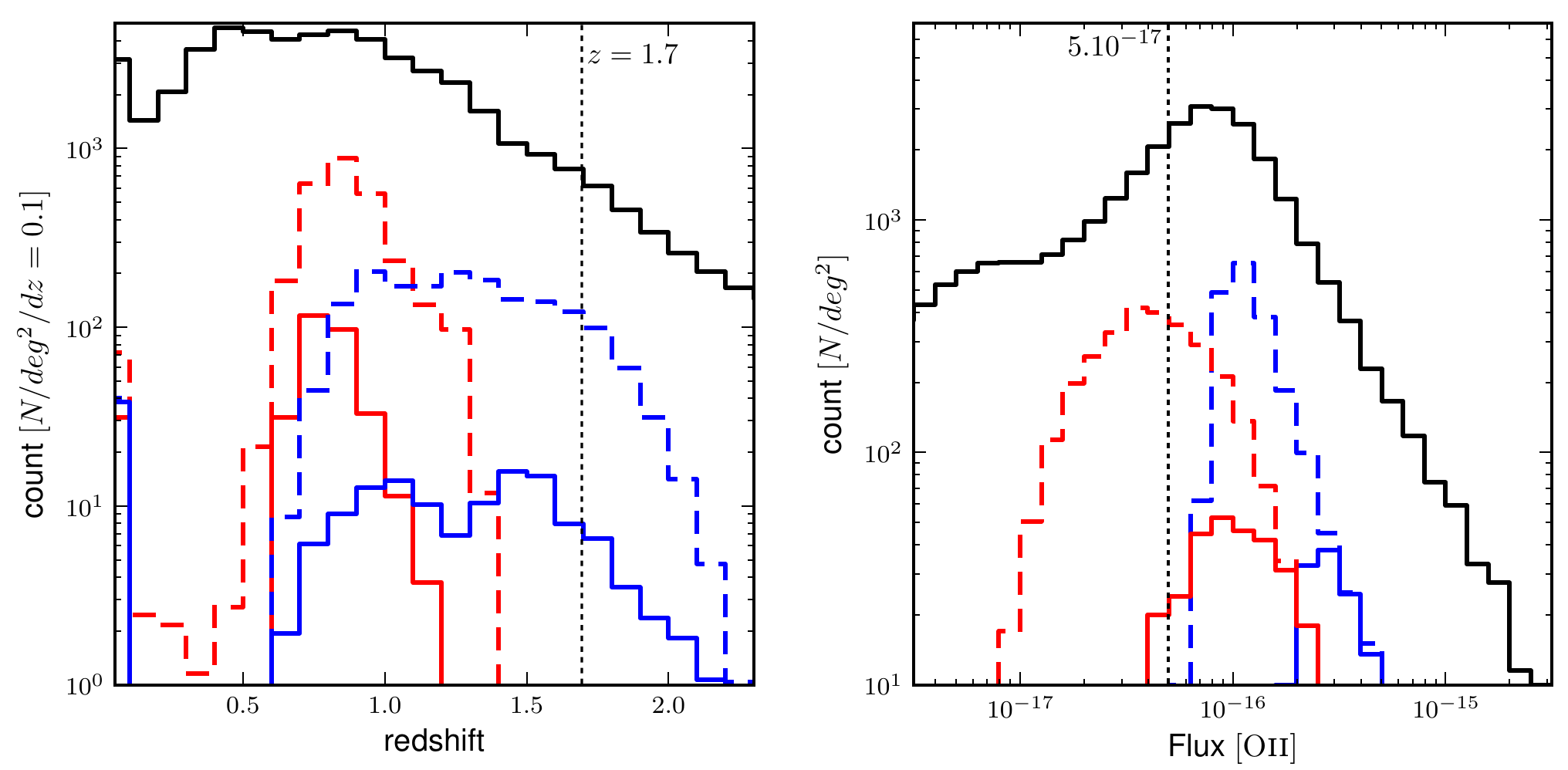}
\caption{Redshift and line flux distribution for various color selections. The blue solid line is the \emph{ugr} selection limited to $20<g<22.5$. The blue dashed line is the \emph{ugr} selection limited to $22.5<g<23.5$. The red solid line is the \emph{gri} selection limited to $19<i<21.3$. The red dashed line is the \emph{gri} selection limited to $21.3<i<23.0$. \textbf{Left.} Photometric redshift distribution from the CFHT-LS catalog. Black solid line: no colour selection, but limited to $i<24$. The \emph{ugr} selection is more spread in redshift than the \emph{gri} selection. The vertical dashed line indicates the upper limit of $z=1.7$ that corresponds to $\left[\mathrm{O\textrm{\textsc{ii}}}\right]$ emission lines at $\sim 1 \mathrm{\mu m}$.
\textbf{Right.} Expected $\left[\mathrm{O\textrm{\textsc{ii}}}\right]$ flux distribution from the CMC simulation. The black solid line is the complete CMC catalog. The \emph{ugr} selection identifies stronger line emitters than the \emph{gri} selection scheme. The vertical dashed line indicates the expected mean sensitivity of BOSS in 1h exposure, $5\times 10^{-17} \mathrm{erg\,s^{-1}\,cm^{-2}}$ .
}
\label{selection_figure_bis}
\end{center}
\end{figure*}

The \emph{ugr} colour selection avoids the stellar sequence, but not the quasar sequence. Hence, the contamination of the \emph{ugr} selection by point-source objects is primarily due to quasars; see Fig. \ref{selection_figure} a) and b). The resulting photometric-redshift distribution as derived from the CFHT-LS photometric redshift catalog has a wide span in redshift, covering $0.6<z<2$ as shown in Fig. \ref{selection_figure_bis}. The distribution is centered at $z=1.3$ for the bright and the faint sample with a scatter of $0.3$ (see Table \ref{mock_selections}). The expected $\left[\mathrm{O\textrm{\textsc{ii}}}\right]$ fluxes are computed from the CMC catalog and are shown in Fig. \ref{selection_figure_bis}. For 90 percent of galaxies in the faint sample, the predicted flux is above $10.6 \times 10^{-17}\mathrm{erg\,cm^{-2}\,s^{-1}}$. The bright sample galaxies show strong emission lines. 

The \emph{gri} selection avoids both the stellar sequence and the quasar sequence; see Fig. \ref{selection_figure} c) and d). Thus the contamination from point-sources should be minimal. Fig. \ref{selection_figure_bis} shows the photometric redshift distribution of the \emph{gri} selection applied to CFHT photometry. The redshifts are centered at $z=0.8$ for the bright and $1.0$ for the faint sample (see Table \ref{mock_selections}). The expected $\left[\mathrm{O\textrm{\textsc{ii}}}\right]$ flux, computed with the CMC catalog, is shown in Fig. \ref{selection_figure_bis}. Emissions are weaker than for the \emph{ugr} selection as expected.

The different selections shown in Fig. \ref{selection_figure} and Fig \ref{selection_figure_bis} are summarized in Table \ref{mock_selections}, which contains the number densities available, mean magnitudes, mean redshifts, and mean $\left[\mathrm{O\textrm{\textsc{ii}}}\right]$ fluxes (when available) of the different samples considered. We have lower densities in the CMC than in the CFHT-LS catalog. This is probably due to cosmic variance as the CMC only covers 2 deg$^2$. 

\begin{figure*}
\begin{center}
\includegraphics[width=180mm]{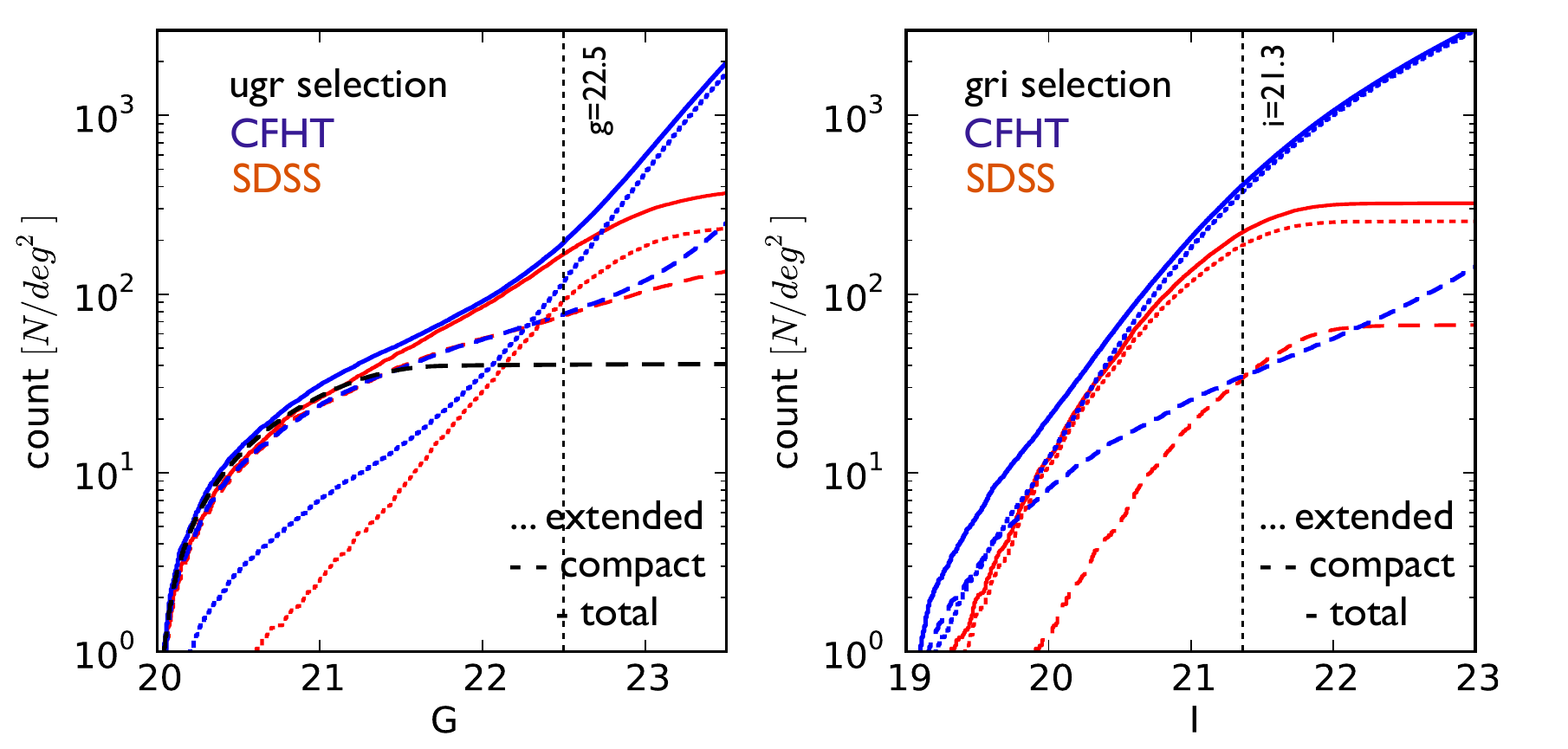}
\caption{Cumulative number counts per square degree of both colour selections using SDSS photometry (in red) and CFHTLS photometry (in blue). The solid line is the whole selection, while the dashed line is for compact (star-like) objects and the dotted line represents extended objects. On the left is the \emph{ugr} selection and on the right is the \emph{gri} selection. The black dashed line represents objects from the XDQSO catalog that have a probability of being a quasar that is greater than 90 percent \citep{Bovy_2011} (the quasar identifications are limited to $g=21.5$).}
\label{cumulative_samples}
\end{center}
\end{figure*}

The SDSS colour-selected samples are complete for the bright samples at $g<22.5$ and $i<21.3$, not for the faint samples. The CFHTLS-selected samples are complete for both the bright and faint samples; see Fig.\ref{cumulative_samples}, where the total cumulative number counts (solid line) of the \emph{ugr} and \emph{gri} colour-selected samples are plotted as a function of  $g$ and $i$ bands respectively. On the bright end of this Figure, although both photometry are complete at the bright limit, we note a discrepancy between the total amount of target selected on CFHT and SDSS that implies selections on CFHT are denser than on SDSS (difference between the red and blue solid lines). This is due to the transposition of the color selection from one photometric system to the other.  In fact, we select targets on SDSS with a transposed criterion from CFHT using the calibrations by \citet{Regnault_2009}. The transposed criterion is as tight as the original. But as the errors on the magnitude are larger in the SDSS system, their colour distributions are more spread. Therefore the SDSS selection is a little less dense than the CFHT selection.

Targeting the bright range is limited by galaxy density, in the best case one can reach 300 targets deg$^{-2}$ and it contains point-sources (stars and quasars) and low-redshift galaxies. In the faint range, the target density is ten times greater, but the exposure time necessary to assign a reliable redshift will be much longer (one magnitude deeper for a continuum-based redshift roughly corresponds to an exposure five times longer). The stellar, quasars and low-redshift contamination is smaller in the faint range. Fig. \ref{selection_figure_bis} shows the distributions in redshift and in $\left[\mathrm{O\textrm{\textsc{ii}}}\right]$ flux we expect given a magnitude range and a colour criterion within the framework of the CMC simulation. The main trend is that the \emph{ugr} selection identifies strong $\left[\mathrm{O\textrm{\textsc{ii}}}\right]$ emitters out to $z\sim2$ where the \emph{gri} peaks at redshift $1$ and extends to $1.4$ with weaker $\left[\mathrm{O\textrm{\textsc{ii}}}\right]$ emitters.

%------------------------- Figure 2bis: ugr, gri, redshift vs OII n(z) expected --------------------------%

We also used a criterion to split targets in terms of compact and extended sources, which is illustrated in Fig. \ref{cumulative_samples}. For CFHT-LS we have used the half-light radius ($r_2$ value, to be compared to the $r_{2}^{limit}$ value which defines the maximal size of the PSF at the location of the object considered - see Coupon et al 2009 and CFHT-LS T0006 release document) to divide the sample into compact and extended objects. For SDSS we used the ``\textsc{type}'' flag, which separates compact (\textsc{type}=6) from extended objects (\textsc{type}=3). For the \emph{ugr} colour selection, the number counts are dominated by compact blue objects (quasars) at $g\leq22.2$. At $g\geq 22.2$ the counts are dominated by extended ELGs. For comparison we show in Fig. \ref{cumulative_samples} the cumulative counts of the XDQSO catalog from \citet{Bovy_2011} who identified quasars in the SDSS limited to $g<21.5$. We notice an excellent match with the bright (compact) \emph{ugr} colour-selected objects. For the \emph{gri} colour selection, there is a low contamination by compact objects because the colour box does not overlap with either the stellar or the quasar sequence.

%These characteristics demonstrate that the \emph{ugr} and \emph{gri} colour boxes can be used to perform an efficient and observable ELG selection.

%------------ Figure 4: cumulative count/sqDeg plot for QSO ELG for CFHT and SDSS ----------------------%

%============== ============= ============= ========== 
%============= ============= ============= =========== 
%=============		Measurements     =============
%=============== ===================  ================
%=============== ===================  ================
\section{ELG Observations}
\label{section:Measurements}

To test the reliability of both the bright \emph{ugr} ($g<22.5$) and the bright \emph{gri} ($i<21.3$) colour selections, we have conducted a set of dedicated observations, as part of the ``Emission Line Galaxy SDSS-III/BOSS ancillary program''. The observations were conducted between Autumn 2010 and Spring 2011 using the SDSS telescope with the BOSS spectrograph at Apache Point Observatory. A total of $\sim$2,000 spectra, observed 4 times 15 minutes, were taken in different fields: namely, in the Stripe 82 (using single epoch SDSS photometry for colour selection) and in the CFHT-LS W1, W3 and W4 wide fields (using CFHT-LS photometry). This data set was released in the SDSS-III Data Release 9\footnote{http://dr9.sdss3.org/}.
 
%\subsection{Description of BOSS spectra}
\subsection{Description of SDSS-III/BOSS spectra}

We used the SDSS photometric catalog \citep{SDSS_DR8} to select 313 objects according to their \emph{ugr} colours located in the Stripe-82 and 899 objects selected according to their \emph{gri} colours in the CFHT-LS W3 field.  In addition we used the CFHT-LS photometry to select 878 \emph{ugr} targets in the CFHT-LS W1 field, and 391 \emph{gri} targets in the CFHT-LS W3 field for observation. The spectra are available in SDSS Data Release 9 and flagged `ELG'.

All of these spectra were manually inspected to confirm or correct the redshifts produced by two different pipelines (\textsc{zCode} and its modified version that we used to fit the $\left[\mathrm{O\textrm{\textsc{ii}}}\right]$ emission line doublet). As the BOSS pipeline redshift measurement is designed to fit LRG continuum some ELG with no continuum were assigned wrong redshifts. To classify the observed objects, we have defined seven sub-categories :

\subsubsection*{Objects with secure redshifts}
\begin{itemize}
\item `ELG', Emission-line galaxy (redshift determined with multiple emission lines). Usually these spectra have a weak `blue' continuum and lack a `red' continuum. Empirically, using \textsc{platefit vimos} pipeline output, this class corresponds to a spectrum with more than two emission lines with observed equivalent widths $EW \leq -6 \AA$ ; see examples in Appendix \ref{tble_appendix}.% \ref{Em_loz}, \ref{Em_midz}, \ref{Em_hiz}.
\item `RG', Red Galaxy with continuum in the red part of its spectrum, allowing a secure redshift measurement through multiple absorption lines ( {\it e.g.} Ca K\&H, Balmer lines) and the $4000\AA$ break.  Some of these objects have also weak emission (E+A galaxies). Empirically their spectra have a mean $D_n(4000)$ of $1.3$ ; where $D_n(4000)$ is the ratio of the continuum level after the break and before the break. These galaxies typically have $i\sim20$, which is fainter than the CMASS targeted by BOSS.
\item `QSO', Quasars, which are identified through multiple broad lines. Examples are given in Fig. \ref{qsos}.
\item Stars.
\end{itemize}

\subsubsection*{Objects with unreliable redshifts}
\begin{figure}
\begin{center}
\includegraphics[width=88mm]{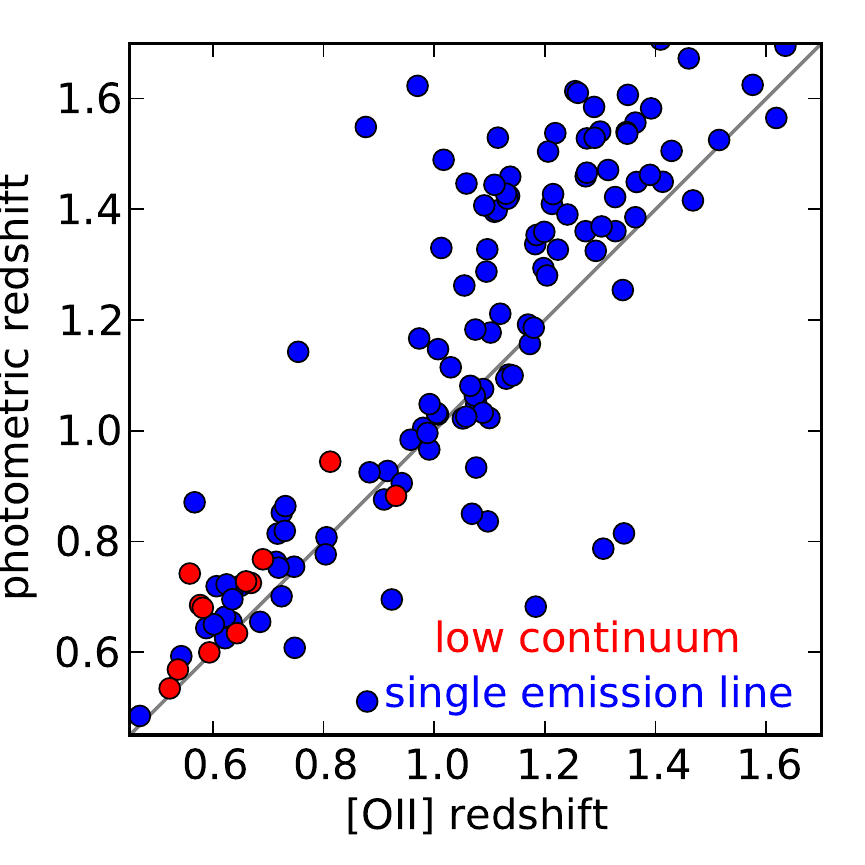}
\caption{T0006 CFHT-LS photometric redshifts of single emission line and low continuum galaxies observed against $\left[\mathrm{O\textrm{\textsc{ii}}}\right]$ redshift. A strong correlation is clearly evident.  A slight systematic over-estimation of the photometric redshift is visible above redshift 1.2 (these photometric redshifts were calibrated below 1.2).}
\label{sinlgeEmLowContiRedshift}
\end{center}
\end{figure}

\begin{itemize}
\item `Single emission line' : the spectra contain only a single emission line which cannot allow a unique redshift determination.
For this population, the CFHT T0006 photometric redshifts are compared to the $\left[\mathrm{O\textrm{\textsc{ii}}}\right]$ redshift (assuming the single emission line is $\left[\mathrm{O\textrm{\textsc{ii}}}\right]$) in Fig. \ref{sinlgeEmLowContiRedshift}. 
The two estimates agree very well : 77.7 percent have $(z_{spec} - z_{phot})/(1+ z_{spec})<0.1$ for the \emph{gri} selection and 62.7 percent for the \emph{ugr} selection.
These galaxies with uncertain redshift tend to have slightly fainter magnitudes with a mean CFHT \emph{g} magnitude at 22.6 and a scatter of 0.6, whereas for the whole ELGs is 22.4 with a scatter of 0.4.
\item `Low continuum' spectra that show a $4000 \AA$ break too weak for a secure redshift estimate. The agreement between photometric and spectroscopic redshift estimation is excellent : 84.6 percent within 10 percent errors; see Fig. \ref{sinlgeEmLowContiRedshift}. 
\item `Bad data', the spectrum is either featureless, extremely noisy or both. %Nothing can be inferred from it. Basically not enough photons went down the fiber.
\end{itemize}
The detailed physical properties of the ELGs are discussed in section \ref{properties} and a number of representative spectra are displayed in Appendix \ref{tble_appendix}.%Figures \ref{Em_loz}, \ref{Em_midz}, \ref{Em_hiz}, \ref{qsos}.

%----------------- Figure 7. photoZ [Oii] spectroZ ----------------------------

%\subsection{Redshift Identification}
\subsection{Redshift Identification}
The results of the observations are summarized by categories in Table \ref{objects_W3}.

For the targets selected using SDSS photometry and with the \emph{ugr} selection : 32 percent are ELGs at a redshift $z>0.6$ (100 spectra). The low-redshift ELGs represent 32 percent of the observed targets (101 spectra).The other categories are : 65 `bad data' (20 percent), 30 quasars (10 percent), 10 stars (3.5 percent), and 7 red galaxies with $z<0.6$ (2.5 percent). With the \emph{gri}-selection, 57 percent of the targets are at $z>0.6$. However, still 21 percent of the spectra fall into the bad data class. 

Using CFHTLS photometry 46 percent of targets are ELGs at $z>0.6$ and 14 percent are quasars with the \emph{ugr} selection. With the \emph{gri}-selection, 73 percent are galaxies at $z>0.6$, five-sixths of which are ELG. 

\begin{figure}
\begin{center}
\includegraphics[width=88mm]{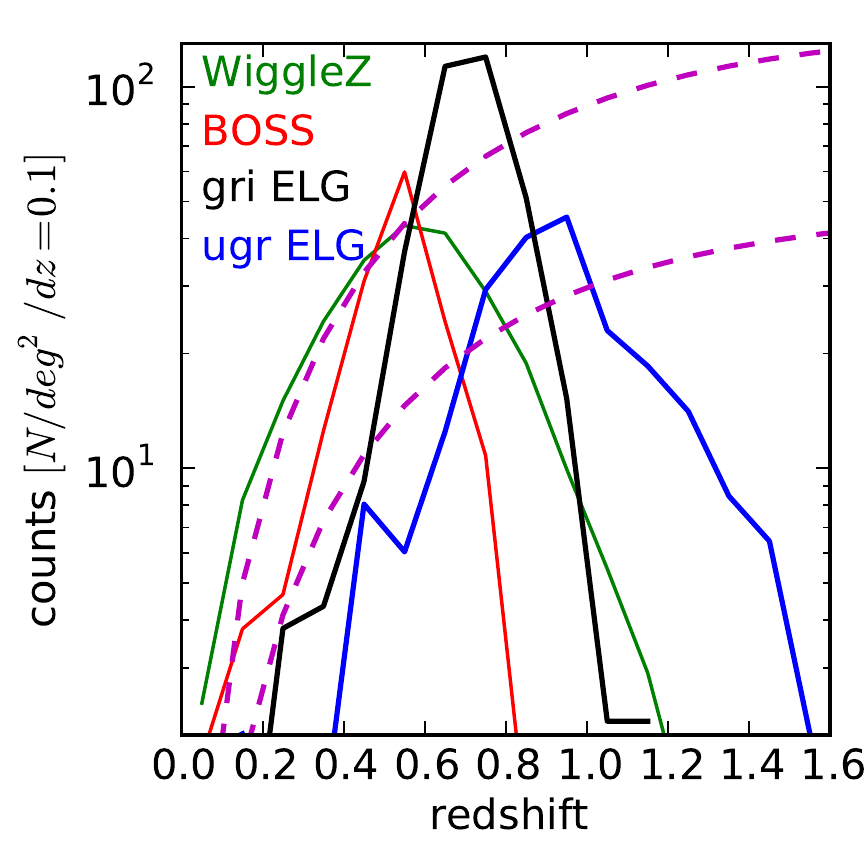}
\caption{Observed redshift distribution for the \emph{ugr} ELGs (blue), the \emph{gri} ELGs (black) compared to the distribution of galaxies from BOSS (red) and WiggleZ (green). Magenta lines represent constant density of galaxies at 1 and 3 $\times10^{-4}\; h^3 \;{\rm Mpc}^{-3}$, it constitutes our density goals.}
\label{ELG_nz}
\end{center}
\end{figure}

For both selections, targeting with CFHTLS is more efficient than with SDSS. The complete classification of observed targets is in Table \ref{objects_W3}. The redshift distribution of the observed objects is compared to the distributions from the BOSS and WiggleZ current BAO experiments in Fig. \ref{ELG_nz}. The Figure shows that \emph{ugr} and \emph{gri} target selections enable a BAO study at higher redshifts. With a joint selection, we can reach the requirements described in Table \ref{BAO_req} to detect BAO feature to redshift 1.

%------------------ Table 2 objects in groups ------------------------------%
\begin{table*}
	\caption{Observed objects split in categories.}
	\label{objects_W3}
	\centering
	\begin{tabular}{l r r r r r r r r }
	\hline \hline	
		&\multicolumn{4}{c}{\emph{gri} selection} & \multicolumn{4}{c}{\emph{ugr} selection} \\	
		& \multicolumn{2}{c}{\emph{SDSS} selection} & \multicolumn{2}{c}{\emph{CFHTLS} selection}& \multicolumn{2}{c}{\emph{SDSS} selection} & \multicolumn{2}{c}{\emph{CFHTLS} selection} \\
		Type & 			Number	&	\% & Number	&	\% & Number	&	\% & Number	&	\%	\\ \hline
		ELG($z>0.6$) & 	450		&	50 	& 	240		&	61  	& 100	& 32  &402	&46\\
		ELG($z<0.6$) &	 60		&	7	&	 3		&	1	& 101 	& 32  & 84 	& 9 \\
		RG($z>0.6$) &	 	73		&	8	&	 46		&	12 	&   0		&  0  	&  0		& 0\\
		RG($z<0.6$) &		 30		&	3	&	 0		&	0	&   7		&  3  	&  0		& 0\\
		single emission line&	 36		&	4  	&	 12		&	3  	&   0		&  0  & 102		& 12\\
		low continuum &	 13		&	1	&	 1		&	0	&   0		&  0  &  0		& 0\\
		QSO & 			 8		&	1   	&	 5		&	1  	&  30		& 10  & 126	&14\\
		stars & 			44		&	5   	&   12		&	3  	&  10		&  3  &  6		& 1\\
		bad data 		& 	185		&	21	& 	72		&	18	&  65 	& 20	  & 158 	& 18\\
		\hline		
		total 			& 	899		&	100	& 391		&	100	& 313 	& 100 & 878 & 100\\ \hline
	\end{tabular}
\end{table*}

%\subsection{Simulation of expected detections with BOSS}
\subsection{Comparison of measured ELGs with the CMC forecasts}

%------------ Figure 4: detections flux OII, sky BOSS, redshift ----------------------%
\begin{figure*}
\begin{center}
\includegraphics[width=180mm]{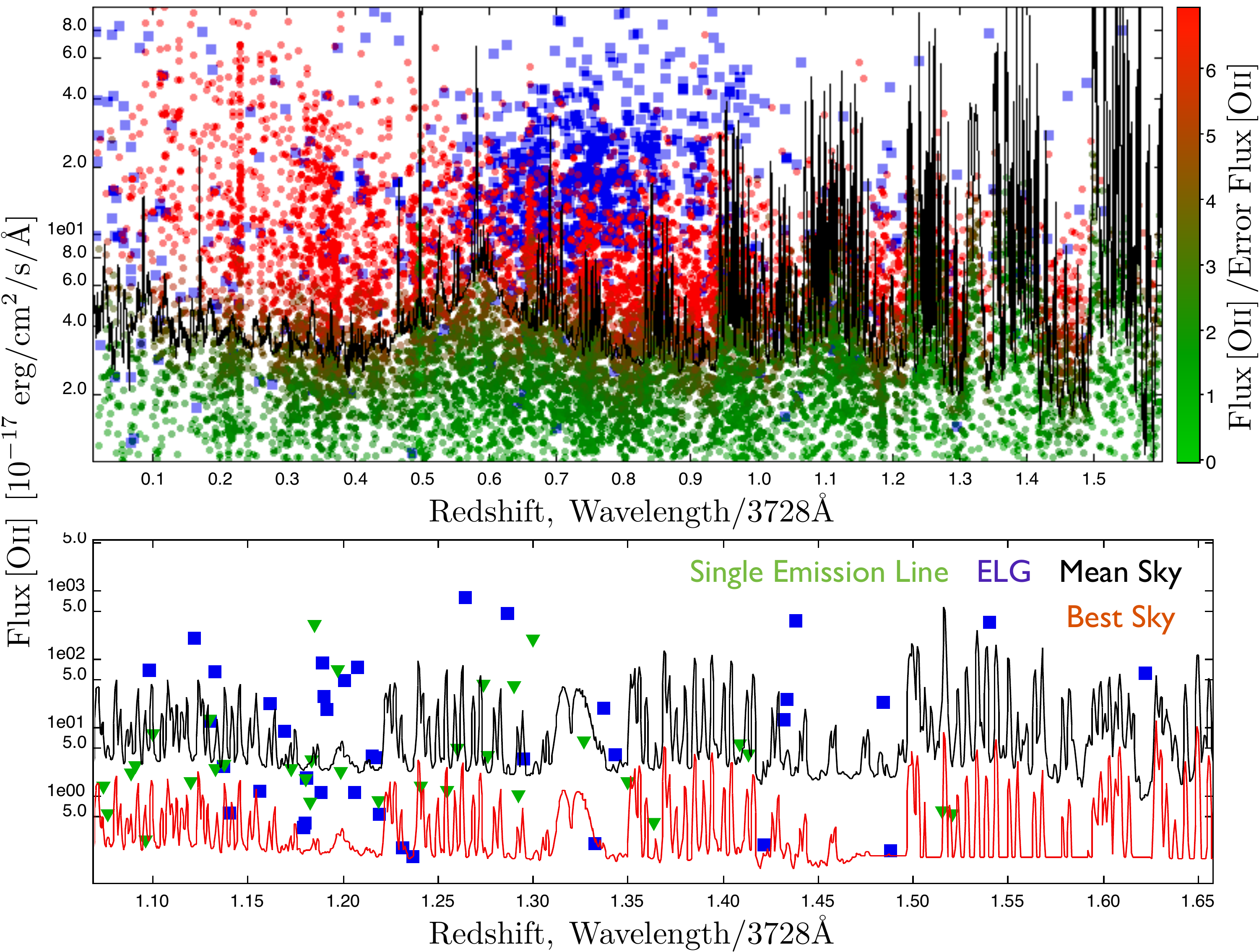}
\caption{Top : Flux detected in the $\left[\mathrm{O\textrm{\textsc{ii}}}\right]$ doublet (green or red filled circle for a simulated spectrum and blue filled square for a real galaxy observation) versus redshift. The simulation is coloured according to the measured line flux divided by the error. The black solid line represents a typical sky as observed by BOSS spectrograph; the $5\sigma$ detections of $\left[\mathrm{O\textrm{\textsc{ii}}}\right]$ follow the sky flux. The $\left[\mathrm{O\textrm{\textsc{ii}}}\right]$ detections above $5\sigma$ (red circles) follow the tendency of the sky fluctuations : a $5\sigma$ detection in a zone with a high sky requires a larger emitted flux. The data are scaled to match a 1 hour exposure on the 2.5m SDSS telescope. The redshifts below the sky flux are assigned using other emission lines or the information in the continuum.
\textbf{Bottom:} : An expansion of the top panel using $z>1$ detections. Only measured galaxies are plotted upon the mean sky (black) and the best sky (red) of APO.
%\textbf{Bottom:} Total flux in $\left[\mathrm{O\textrm{\textsc{ii}}}\right]$ emission lines against the median sky flux measured between the wavelengths between $3727(1+z)$ and $3729(1+z)$. One panel per modified julian date (MJD). The crosses represent all observations of the day through their median (center of the cross) and their standard deviation (arms of the cross). They split in three groups. Good detections with $z>1.1$ in green. Single emission lines that would be at $z>1.1$ if $\left[\mathrm{O\textrm{\textsc{ii}}}\right]$ is the emission line in orange. Lousy detections that would be at $z>1.1$ according to the photometric redshift of the target in red.
}
\label{OII_detection_limit}
\end{center}
\end{figure*}

To investigate the expected purity of ELG galaxies samples, we created mock catalogs covering redshifts between 0.6 and 1.7. Continuum spectra of ELGs were generated from the Cosmos Mock Catalog and emission lines were added according to the modeling described in \citet{Jouvel_2009}. Two simulated galaxy catalogs were built, one for each colour selection function (\emph{ugr} and \emph{gri}). Each synthetic spectrum was affected by sky and photon noise as if observed by BOSS spectrographs, by using the \textsc{specsim1d} software. We simulated a set of four exposures of 900 seconds each. The resulting simulated spectra were then analyzed by the \textsc{zCode} pipeline \citep{2006MNRAS.372..425C} to extract the spectroscopic redshift. As our targets are mainly emission line galaxies, we only use the redshift estimate based on fitting discrete emission line templates in Fourier space over all z.

We address the flux measurement of emission lines. This exercise was conducted using the \textsc{Platefit Vimos} software developed by \citet{Lamareille_2009}. %Lamareille et al. 2009.
%Here we derive the sensitivity of the redshift solver to $\left[\mathrm{O\textrm{\textsc{ii}}}\right]$ line strength and the redshift.
%For this, we create mock observations of the \emph{ugr} and the \emph{gri} galaxy samples with the SDSS spectrograph. For this we use the \textsc{SpecSim1d} software. The output of the simulation is a one dimensional spectrum as observed with the \emph{BOSS} spectrographs. We simulate a set of four exposures of 900 seconds each. Spectroscopic redshifts for each simulated spectra are fit using the \textsc{zCode} pipeline \citep{2006MNRAS.372..425C}. As our targets are mainly emission line galaxies, we only use the redshift estimate based on fitting discrete emission line templates in Fourier space over all z. 
This software is based on the \textsc{platefit} software that was developed to analyze %high resolution 
SDSS spectra \citep{2004ApJ...613..898T,2004MNRAS.351.1151B}. The \textsc{platefit vimos} software was developed to measure the flux of all emission lines after removing the stellar continuum and absorption lines from lower resolution and lower signal-to-noise ratio spectra \citep{2006A&A...448..907L}. The stellar component of each spectrum is fit by a non-negative linear combination of 30 single stellar population templates with different ages (0.005, 0.025, 0.10, 0.29, 0.64, 0.90, 1.4, 2.5, 5 and 11 Gyr) and metallicities (0.2, 1 and 2.5 $Z_\odot$). These templates have been derived using the \citet{2003MNRAS.344.1000B} libraries and have been resampled to the velocity dispersion of VVDS spectra. The dust attenuation in the stellar population model is left as a free parameter. Foreground dust attenuation from the Milky Way has been corrected using the \citet{1998ApJ...500..525S} maps. 

After removal of the stellar component, the emission lines are fit as a single nebular spectrum made of a sum of Gaussians at specified wavelengths. All emission lines are set to have the same width, with the exception of the $\left[\mathrm{O\textrm{\textsc{ii}}}\right]\lambda3727$ line which is a doublet of two lines at 3726 and 3729 $\AA$ that appear broadened compared to other single lines.
Detected emission lines may also be removed from the original spectrum in order to obtain the observed stellar spectrum and measure indices, as well as emission-line equivalent widths. The underlying continuum is obtained by smoothing the stellar spectrum. Equivalent widths are then measured via direct integration over a $5\sigma$ bandpass of the emission-line Gaussian model divided by the underlying continuum. Then emission lines fluxes are measured for each simulated spectra using the extracted redshift from \textsc{zCode} and the true redshift for cross-checks.

We consider that a redshift has been successfully measured if $\Delta z/(1+z)<0.001$. We believe that this threshold could be lowered to $10^{-4}$ in the future by using a more advanced redshift solver. Using the current pipeline, we can distinguish these two regimes.

The first regime is the redshift range $z<1.0$. Many emission lines ([OII], H$\beta$, [OIII]) are present in the SDSS spectrum. For $g<23.5$, 91 percent of the redshift are measured sucessfully. Among the remaining 9 percent, catastrophic failures represent 3.5 percent (the pipeline outputs a redshift between 0 and 1.6 with $\Delta z/(1+z)>0.01$). Inaccurate redshifts represent 3.9 percent (the pipeline outputs a redshift between 0 and 1.6 with $0.001<\Delta z/(1+z)<0.01$) and 1.5 percent are not found by the pipeline ($z=-9$ is output).

The second regime is the redshift range $1.0\leq z< 1.7$: the redshift determination hinges on the identification of the $\left[\mathrm{O\textrm{\textsc{ii}}}\right]$ doublet. For $g<23.5$, 66.8 percent of the redshifts are measured sucessfully. 19.1 percent are catastrophic failures and 14.1 percent are inaccurate redshifts. Work is ongoing to improve the redshift measurement efficiency at $z>1$. In the second regime, the minimum $\left[\mathrm{O\textrm{\textsc{ii}}}\right]$ flux required to compute a reliable redshift depends on the redshift / wavelength, because of the strong OH sky lines in the spectrum. We infer from the observed spectra that to measure a reliable redshift, we require a $5\sigma$ detection of the $\left[\mathrm{O\textrm{\textsc{ii}}}\right]$ lines, which means a (blended or not) detection of two peaks in the emission line separated by $2(1+z)$. The detection significance is defined from the 1d spectrum. From the data the faintest $5\sigma$ detections are made with a flux of $4\times 10^{-17} \mathrm{erg\,s^{-1}\,cm^{-2}}$ and the brightest $5\sigma$ detections need a flux of $2\times 10^{-16} \mathrm{erg\,s^{-1}\,cm^{-2}}$ to be on top of sky lines. The simulation shows the same thresholds; see Fig. \ref{OII_detection_limit}. The simulation confirms the detection limit we observe. The bottom plot of Fig. \ref{OII_detection_limit} raises the issue that the sky time variation has a non-negligible impact on the detection limit of the $\left[\mathrm{O\textrm{\textsc{ii}}}\right]$ emission doublet for redshifts $z>1.1$. Though this ELG sample is too small to address this issue. In fact the sample was observed during ten different nights and the number of ELG with $z>1.1$ is less than 60. It is thus not possible to derive a robust trend comparing the $\left[\mathrm{O\textrm{\textsc{ii}}}\right]$ detections to the sky value of each observation. Handling this issue would require a sample of $\sim$500 redshifts in $1.1<z<1.6$ observed many times over many nights. With such a sample in hands, we could quantify exactly how to optimize the observational strategy.

%============== ============= ============= ========== 
%============= ============= ============= =========== 
%=============	  Physical properties of ELGs    =============
%=============== ===================  ================
%=============== ===================  ================
\section{Physical properties of ELGs}
\label{properties}
All \emph{ugr} and \emph{gri} ELG spectra were analyzed with two different software packages: the PlateFit VIMOS \citep{Lamareille_2009} and the Portsmouth Spectroscopic Pipeline \citep{2012arXiv1207.6115T}. In this section we discuss the following physical properties of the observed ELGs: redshift, star forming rate (SFR), stellar mass, metallicity and classification of the ELG type (Seyfert 2, LINERs, SFG, composite). Observations a larger samples of ELGs are planned to estimate how these quantities vary over time and with their environment, and also to estimate how the clustering depends on these physical quantities. It is key to replace future BAO tracers in the galaxy formation history. With the current sample, we draw simple trends using means and standard deviation of the observed quantities, and we place the ELGs in the galaxy classification made by \citet{Lamareille_2010,Marocco_2011}.

%\subsection{Main Properties}
\subsection{Main Properties}
The main properties of the ELGs are shown in the Table \ref{main_properties}. The star forming rate was computed using the equation 18 of \citet{Argence_2009}. The stellar mass was estimated using the CFHTLS \emph{ugriz} photometry. (The errors on the stellar mass using only SDSS photometry were too large to be meaningful, thus the empty cells in the table). The metallicity is estimated using the calibration by \citet{Tremonti_2004}. The main trends are :
\begin{itemize}
\item The \emph{gri}-selected galaxies of CFHTLS are the more massive galaxies in terms of stellar mass. 
\item The \emph{ugr} selects stronger star-forming galaxies than \emph{gri} (due to the \emph{u}-band selection). There is a factor of two variations in the strength of the measured oxygen lines.
\item The \emph{ugr} selects galaxies that have $12 + \log{[OH]} \in [8,9]$ whereas \emph{gri} focuses slightly more on the higher $12 + \log{[OH]} \approx 9$.
\item the SFR appears to be independent of the color selection schemes.
\end{itemize}
%------------------------- Table A.1: Properties of ELGs --------------------------%
\begin{table*}
	\caption{Main properties of the observed galaxies. Fluxes are in unit of $10^{-17} \,\mathrm{erg \,cm^{-2}\,s^{-1}}$. Equivalent Widths `EW' are in $\AA$.}
	\label{main_properties}
	\centering
	\begin{tabular}{c cc cc cc cc}
	\hline \hline
	 & \multicolumn{4}{c}{{\it gri}-selected ELG} & \multicolumn{4}{c}{{\it ugr}-selected ELG} \\
	 & \multicolumn{2}{c}{CFHTLS} & \multicolumn{2}{c}{SDSS} & \multicolumn{2}{c}{CFHTLS} & \multicolumn{2}{c}{SDSS} \\
	 & mean & $\sigma$ & mean & $\sigma$ & mean & $\sigma$ & mean & $\sigma$ \\
	\hline
	EW$_{\left[\mathrm{O\textrm{\textsc{ii}}}\right]}$  	&   -14.86 & 9.01    	&-16.75 & 10.13	&  -50.58 & 27.24&-30.75 & 23.04\\
	Flux$_{\left[\mathrm{O\textrm{\textsc{ii}}}\right]}$	&    16.85 & 9.65    	&  18.58 & 10.37	&30.36 & 30.1& 24.23 & 39.27\\
	EW$_{H_\beta}$ 							&     -10.28 & 10.8  	&-10.72 & 8.65		&   -24.27 & 22.88&  -17.18 & 19.34\\\
	Flux$_{H_\beta}$ 							&     15.44 & 8.6	&  14.63 & 7.72		&  12.97 & 15.16&12.57 & 23.91\\
	EW$_{\left[\mathrm{O\textrm{\textsc{iii}}}\right]}$ 	&   -10.09 & 10.98	& -11.33 & 10.76	&  -65.3 & 91.56& -16.89 & 30.49\\
	Flux$_{\left[\mathrm{O\textrm{\textsc{iii}}}\right]}$ 	&  17.74 & 20.15	& 17.43 & 21.59	&   35.13 & 53.49&  13.39 & 37.79\\
	$12 + \log OH$ 							&   8.94 & 0.20		&8.92 & 0.19		&  8.69 & 0.21 &  8.69 & 0.25\\
	$\log$SFR$_{\left[\mathrm{O\textrm{\textsc{ii}}}\right]}$ 	&   0.97 & 0.35	&0.92 & 0.45		&  0.96 & 1.24 &  0.76 & 0.84\\
	$\log(M^*/M_\odot)$ 							&   10.85 & 0.3		&10.23 & 6.87		& 9.33 & 0.80 &   - & -\\
	\hline
	\end{tabular}
\end{table*}

%\subsection{Classification.}
\subsection{Classification.}
%------------------------- Figure A.1: Classification of ELGs ----------------
\begin{figure*}
\begin{center}
\includegraphics[width=180mm]{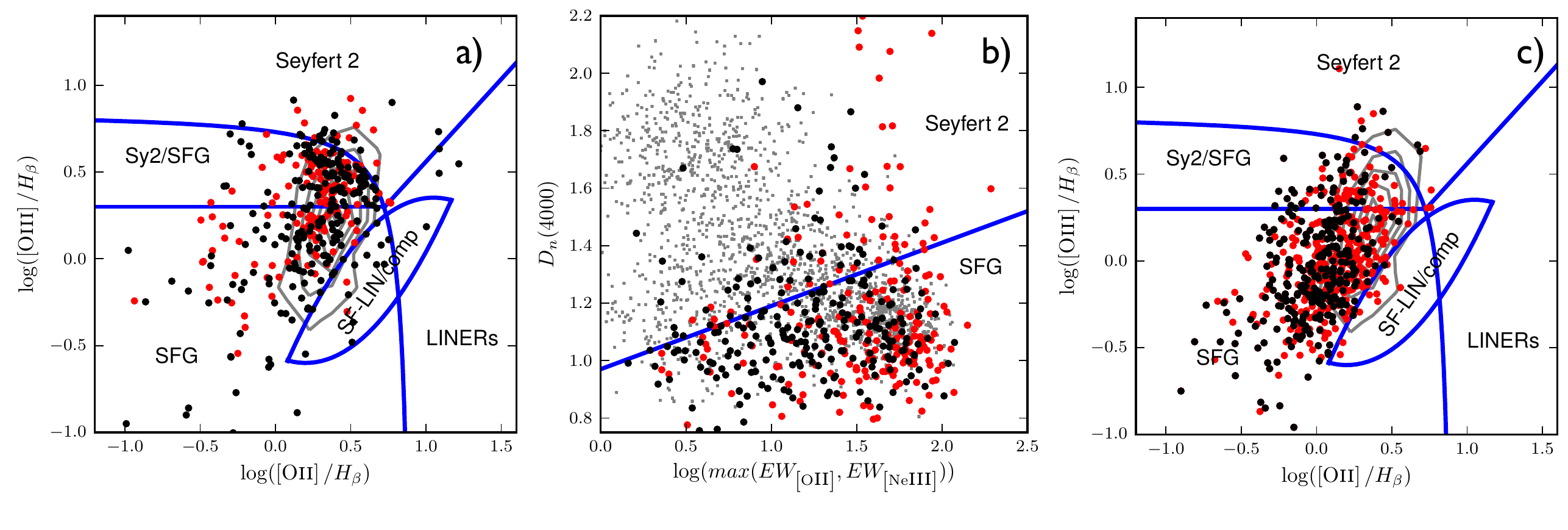}
\caption{Black dots stands for CFHTLS selected ELGs, red dots for SDSS selected ELGs and grey contours or grey pixels for zCOSMOS survey galaxies.
\textbf{a)} $\log(\left[\mathrm{O\textrm{\textsc{iii}}}\right]/H_\beta)$ vs. $\log(\left[\mathrm{O\textrm{\textsc{ii}}}\right]/H_\beta)$ for \emph{ugr} ELGs. The \emph{ugr} ELG sample is located in a similar area than zCOSMOS galaxies. 
\textbf{b)} $D_n(4000)$ vs. $\log(max(EW_{\left[\mathrm{O\textrm{\textsc{ii}}}\right]},EW_{\left[\mathrm{Ne\textrm{\textsc{iii}}}\right]}))$ for \emph{ugr} ELGs. Only galaxies from the `Sy2/SFG' area from the plot a) are represented. The \emph{ugr} ELG is thus mainly composed of `SFG'. 
\textbf{c)} $\log(\left[\mathrm{O\textrm{\textsc{iii}}}\right]/H_\beta)$ vs. $\log(\left[\mathrm{O\textrm{\textsc{ii}}}\right]/H_\beta)$ for \emph{gri} ELGs. The \emph{gri} ELG sample is located in the `SFG' area.}
\label{classification_ELG}
\end{center}
\end{figure*}

We use a recent classification \citep{Lamareille_2010,Marocco_2011} for the ELG sample. The classification is made using $\log(\left[\mathrm{O\textrm{\textsc{iii}}}\right]/H_\beta)$, $\log(\left[\mathrm{O\textrm{\textsc{ii}}}\right]/H_\beta)$, $D_n(4000)$, and $\log(max(EW_{\left[\mathrm{O\textrm{\textsc{ii}}}\right]},EW_{\left[\mathrm{Ne\textrm{\textsc{iii}}}\right]}))$. We compare the ELG sample to zCOSMOS, as zCOSMOS has numerous star forming galaxies in the redshift range we are observing. Fig. \ref{classification_ELG} a) shows that the zCOSMOS and the \emph{ugr} ELG samples are located in three of the five areas delimited by the classification: Seyfert 2 (`Sy2'), Star Forming Galaxies (`SFG'), and a third region where both mix (`Sy2/SFG'). There are a few LINERs and Composite in either sample. Fig. \ref{classification_ELG} b) separates the \emph{ugr} galaxies in the `Sy2/SFG' area into `SFG' or `Sy2', and shows that zCOSMOS galaxies from the `Sy2/SFG' area are both `Sy2' and `SFG' where the \emph{ugr} ELGs in the `Sy2/SFG' area are mostly `SFG'. The \emph{gri} observed sample is located in the area of Star Forming Galaxies (`SFG'), whether one considers the one selected on CFHT or on SDSS. Finally, the ELG selected, \emph{ugr} or \emph{gri}, are both in the `SFG' part of the classification.

%============== ============= ============= ========== 
%============= ============= ============= =========== 
%================	 Discussion		=================
%=============== ===================  ================
\section{Discussion}
\label{section:discussion}

%\subsection{Redshift identification rates in ugr and gri}
\subsection{Redshift identification rates in {\it ugr} and {\it gri}}
We summarize the redshift measurement efficiency of the \emph{gri} and \emph{ugr} colour-selected galaxies presented in this paper in Tables \ref{objects_W3} and \ref{redshift_efficiency}, and we compare the results with those of WiggleZ \citep{Drinkwater_2010}, BOSS and VIPERS (the percentages about VIPERS are based on a preliminary subset including only $\sim$ 20 percent of the survey). The original VIPERS selection flag (J. Coupon and O. Ilbert private communication) is defined to have colours compatible with an object at $z > 0.5$ if it has ($r-i \geq 0.7$ and $u-g\geq1.4$) or ($r-i \geq 0.5(u-g)$ and $u-g<1.4$) (Guzzo et al. (2012), in preparation). The efficiencies in the Table \ref{redshift_efficiency} show that a better photometry and thus more precise colours yield a better efficiency in terms of obtaining objects in the targeted redshift range. It also shows the colour selections proposed in this paper are competitive for building an LSS sample.

\begin{figure}
\begin{center}
\includegraphics[width=88mm]{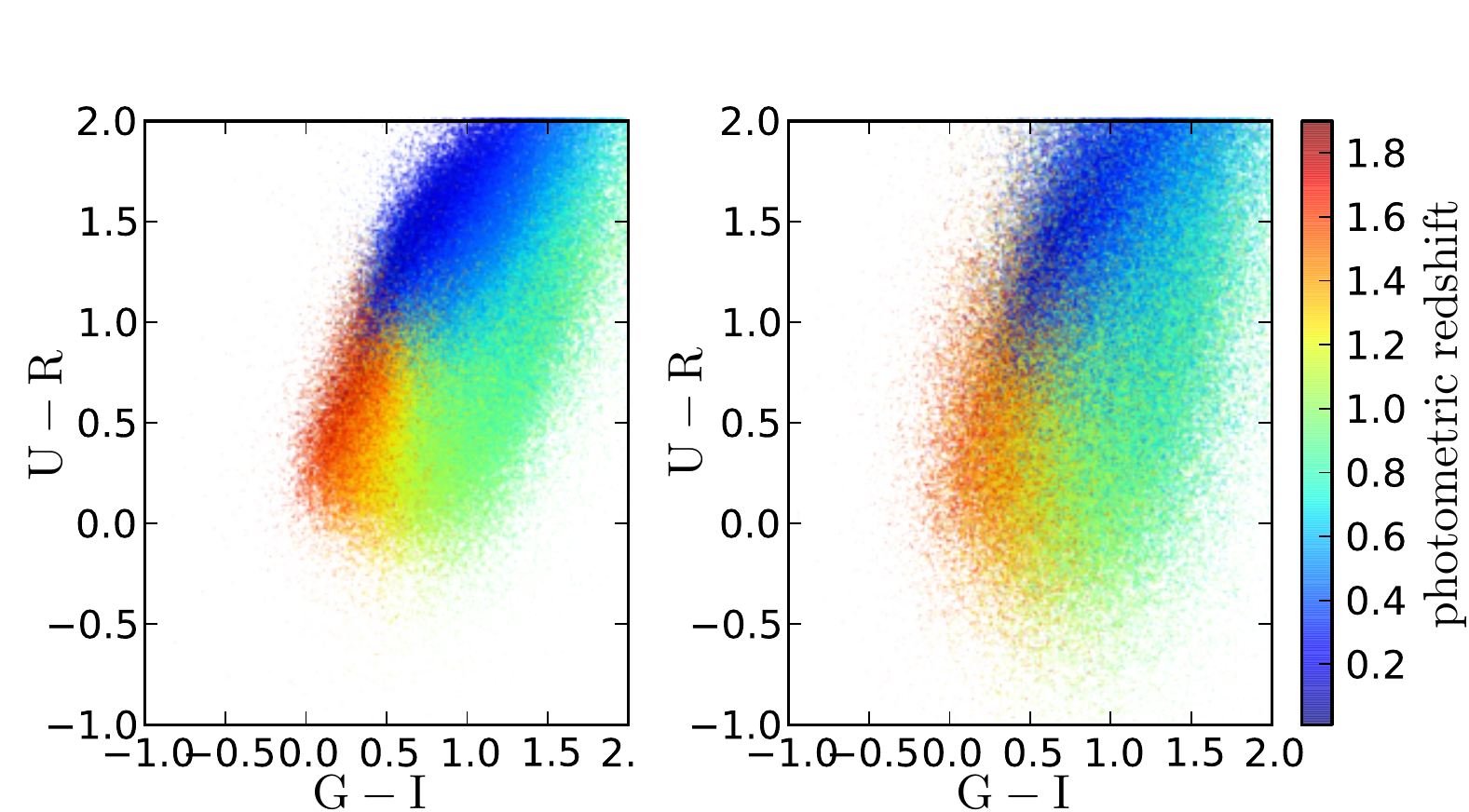}
\caption{U-R vs. G-I colored according to the photometric redshift. On the left CFHT-LS photometry, on the right CFHT-LS photometry degraded by a factor 2.5. This comparison shows how the degradation of the photometry smears the clean separations between galaxy populations in redshift.}
\label{degradedPhotometry:fig}
\end{center}
\end{figure}

To determine the necessary precision on the photometry to stay at the efficiencies observed, we degrade the photometry of the observed ELGs, then reselect them and recompute the efficiencies. Using a photometry less precise than the CFHTLS by a factor of 2.5 in the errors (the ratio of the median values of the mag errors in bins of 0.1 in magnitude equals 2.5) does not significantly change neither the efficiency nor the redshift distribution implied by the colour selection. This change also corresponds to loosening the colour criterion by 0.1 mag. For the \emph{eBOSS} survey a photometry 2.5 times less precise than CFHTLS should be sufficient to maintain a high targeting efficiency (for comparison, SDSS is 10 times less precise than CFHTLS); Fig. \ref{degradedPhotometry:fig} shows the smearing of the galaxy positions in the colour-colour plane for a degraded photometry.

%------------------ Table 3 redshift efficiency ------------------------------%
\begin{table}
	\caption{Redshift efficiency in percent. The second column `spectroscopic redshift' quantifies the amount of spectroscopic redshift obtained with the selection. The third column `object in z window' is the number of spectroscopic redshift that are in the  range the survey is aiming at, it is the efficiency of the target selection. The redshift window for ELG selection is $z>0.6$.}% \textcolor{red}{Not fair! Vipers is very deep over a small field. Either remove or explain in detail. DEEP2 ???}}
	\label{redshift_efficiency}
	\centering
	\begin{tabular}{c c c c c c c c}
	\hline \hline		
		selection		& spectroscopic	&object in  &quasars  \\ 
		scheme		& redshift			&z window & \\
		\hline
		\emph{gri} SDSS 	& 80 		& 62		&1\\
		\emph{gri} CFHTLS	& 82		& 73		& 1 \\
		\emph{ugr} SDSS	& 80 		& 32 		& 10 \\
		\emph{ugr} CFHTLS	& 78 		& 56		& 13 \\
		WiggleZ			& 60 		&  35		& - \\
		BOSS 			& 95 & 95 & - \\
		VIPERS 			&  80 & 70 & - \\
		 \hline
	\end{tabular}
\end{table}
%----------------- Figure 8. gri selection SDSS CFHT ----------------------------

\subsection{Measurement of the $\left[\mathrm{O\textrm{\textsc{ii}}}\right]$ doublet, single emission line spectra}

\begin{figure*}
\begin{center}
\includegraphics[width=180mm]{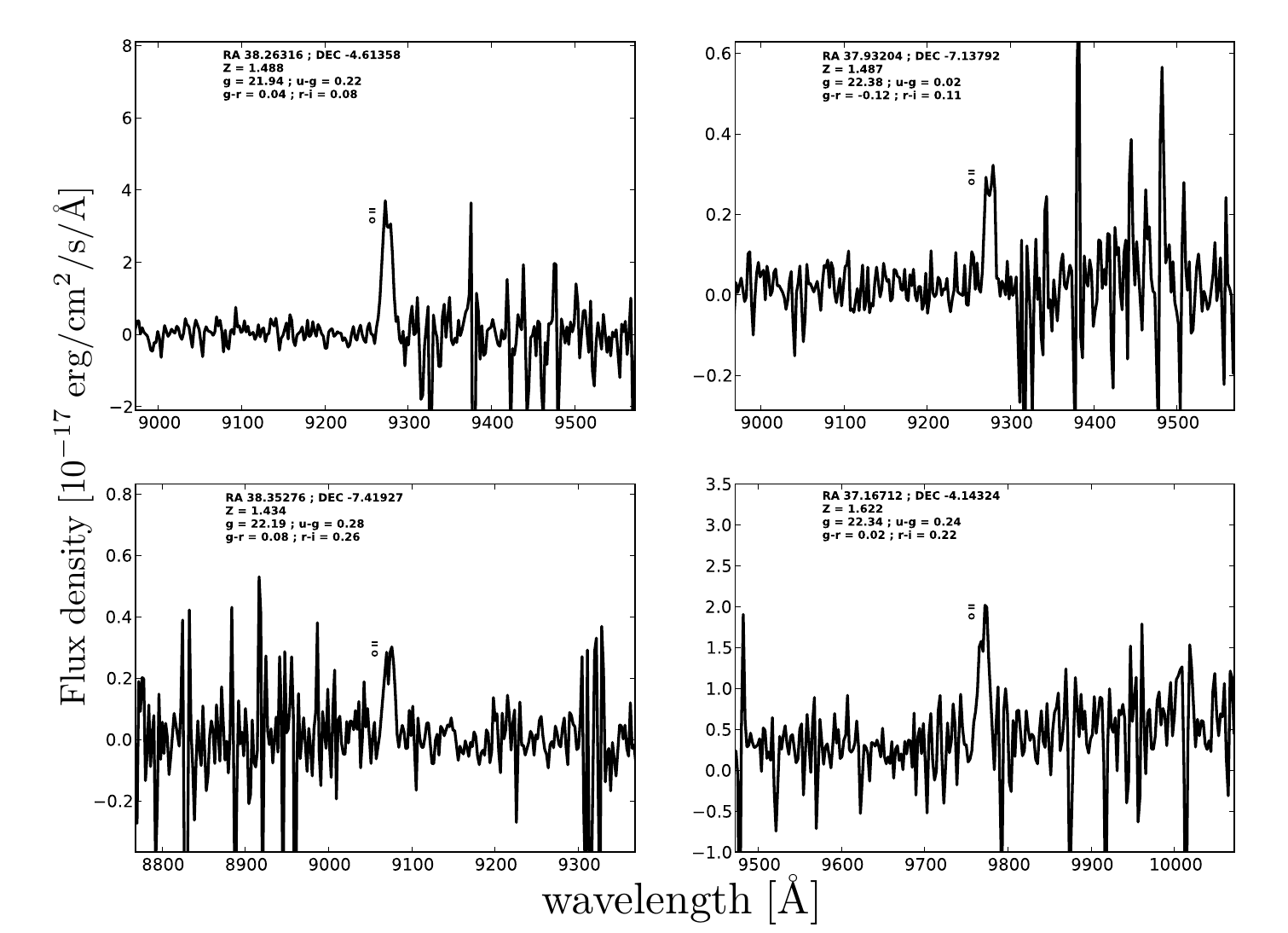}
\caption{Observed Spectra zoomed on the $\left[\mathrm{O\textrm{\textsc{ii}}}\right]$ doublet, extracted from the \emph{ugr} ELG sample. The doublet is enough split (apart from a little blending) to assign the correct spectroscopic redshift without the help of other emission lines.}
\label{OiiRedshifts:Fig}
\end{center}
\end{figure*}

For ground-based spectroscopic surveys observing ELGs with $1<z<1.7$, the only emission line remaining in the spectrum to assign the spectroscopic redshift is the $\left[\mathrm{O\textrm{\textsc{ii}}}\right]$ doublet. For the redshift to be certain the doublet must be split ({\it i.e.}, we do not want the target to be classified as `single emission line' ELG). Fig. \ref{OiiRedshifts:Fig} shows a subsample of the observed bright \emph{ugr} ELGs where $\left[\mathrm{O\textrm{\textsc{ii}}}\right]$ doublets are well resolved.

%----------------- Figure 9.Oii doublets----------------------------

We can circumvent the `single emission line' ELG issue (Fig. \ref{sinlgeEmLowContiRedshift}) by increasing the resolution of the spectrograph. This modification will enhance a better split of $\left[\mathrm{O\textrm{\textsc{ii}}}\right]$, and will increase the room available to observe the doublet by rendering sky lines `thiner'. The sky acts as an observational window and prevents some narrow redshift ranges to be sampled by the spectrograph; see Fig. \ref{OII_detection_limit}. Increasing the resolution dilutes the signal, and thus the exposure time has to be increased to reconstruct properly the doublet above the mean sky level. 

We performed a simulation of the $\left[\mathrm{O\textrm{\textsc{ii}}}\right]$ emission line fit to quantify by which amount the resolution must be increased to have no `single emission line' ELGs. We fit one or two Gaussians on a doublet with a total flux of $10^{-16} \mathrm{erg\,s^{-1}\,cm^{-2}}$ (lowest `single emission line' flux observed) contaminated by a noise of $3 \times 10^{-17} \mathrm{erg\,s^{-1}\,cm^{-2}}$ (typical BOSS dark sky). The $\chi^{2}$ of the two fits are equal at low resolution and become disjoint in favor of the fit with 2 Gaussians for a resolution above 3000 at $7454.2\AA$ ({\it i.e.}  $\left[\mathrm{O\textrm{\textsc{ii}}}\right]$ at redshift 1). Such an increase in resolution could help assigning proper redshifts to `single emission line' ELGs.

\subsection{How / why redshift went wrong}

\begin{figure}
\begin{center}
\includegraphics[width=80mm]{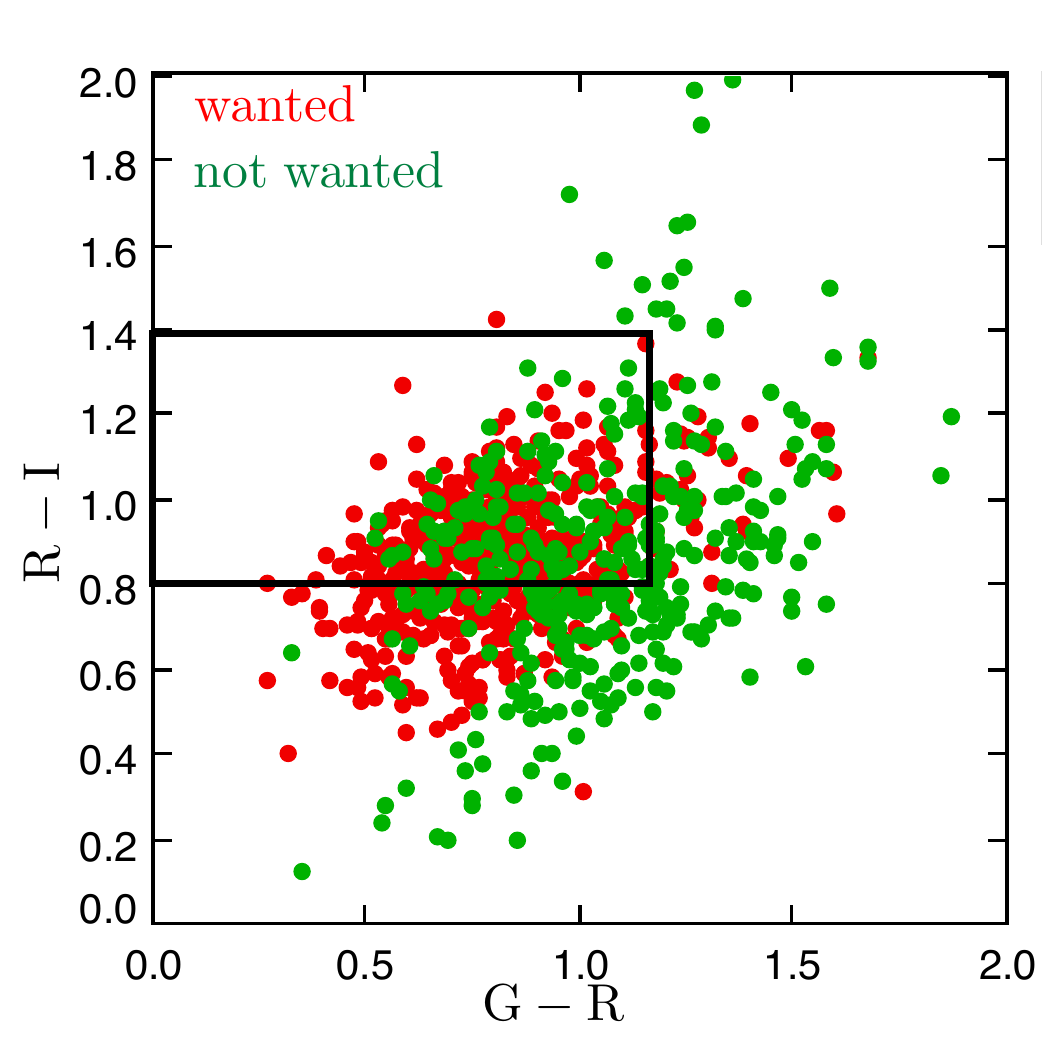}
\caption{\emph{gri} selection based on colours from SDSS (black box) represented on CFHTLS magnitudes. The scatter is quite large: about half the targets would not have been selected if we used CFHTLS photometry. The `wanted' objects are galaxies at $z>0.6$ or quasars and `unwanted' objects are the rest. %Thus as shown with the numbers of Table \ref{objects_W3} a more precise photometry yields a better rejection of unwanted objects.} %\textcolor{red}{What about the "wanted" objects that are not in the colour box?}
}
\label{griComparison}
\end{center}
\end{figure}

The main difference in redshift measurement efficiency between SDSS and CFHT-LS colour selection is mainly due to the difference in photometry depth. Using calibrations made by \citet{Regnault_2009}, it is possible to translate the colour selection criteria from CFHT-LS magnitudes to SDSS magnitudes. The colour difference can be as large as 1 magnitude as the SDSS magnitude cut is close to the detection limit of the SDSS survey; see Fig. \ref{griComparison} where SDSS \emph{gri} colour-selected galaxies are represented with their CFHTLS magnitudes.
%----------------- Figure 8. gri selection SDSS CFHT ----------------------------

%\subsection{How to improve things}
\subsection{How to improve ELG selection for future surveys}
We suggest a few ways to increase the redshift measurement efficiency and reach the requirements set in the second section.

For the \emph{ugr} selection : lowering \emph{u-g} cut to 0.3 diminishes the contamination by low-redshifts galaxies. Additional low-redshift galaxies can be removed from the selection through an inspection of the images. Some of the low-redshift galaxies are quite extended, and one could mistake a high-redshift merger for an extended low-redshift galaxy. Visual inspection reduces the low-redshift share from 9 percent to 4 percent. The compact and extended selection on the CFHT data is very efficient at identifying quasars. There is also room for improving the spectroscopic redshift determination and thus re-classify `single emission line' galaxies : they represent a 12 percent share, among which 10 percent are at $z>0.6$. It seems reasonable to assume an efficiency improvement from 46 percent ELG($z>0.6$) + 14 percent quasar to 61 percent ELG($z>0.6$) + 14 percent quasar. Thus a total efficiency of $\sim75\%$

For the \emph{gri} selection : improving the spectroscopic redshift determination pipeline can gain up to 5 percent efficiency thus increasing from $73$ to 78 percent of  ELG($z>0.6$).

\begin{figure}
\begin{center}
\includegraphics[width=88mm]{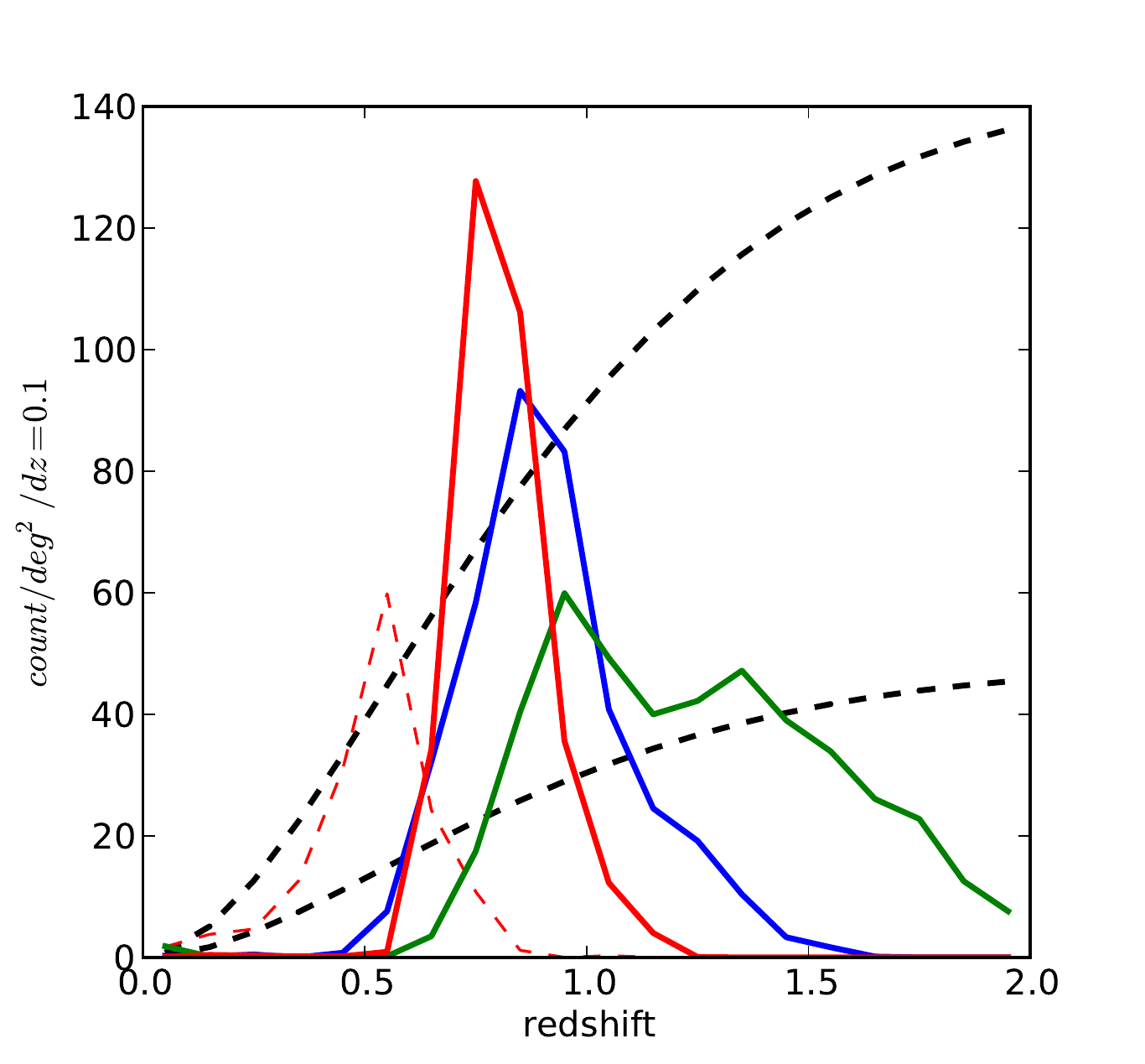}
\caption{Photometric redshift distributions obtained using the \emph{ugri} bands. The dashed black lines are the low and high density goals mentioned in Section \ref{section:ELGs_BAO}, $\bar{n}=10^{-4}$ and $3 \times10^{-4}\; h^3\mathrm{Mpc}^{-3}$. The dashed red line is the BOSS CMASS sample. The solid blue line is the distribution enhanced by the \emph{ugri} selection, it has a projected sky density of $\sim340$ deg$^{-2}$. The solid red line is the \emph{gri} selection (projected sky density $\sim350$ deg$^{-2}$). The solid green is the \emph{ugr} selection (projected sky density $\sim400$ deg$^{-2}$). It shows the possibility of making a selection able to sample $[0.6,1.2]$ for a BAO experiment.}
\label{ugriSelections}
\end{center}
\end{figure}

We have also optimized target selections for BAO sampling density using the four bands \emph{ugri}. We find that the optimum selections have a redshift distribution close to the smooth combination of the \emph{gri} and \emph{ugr} selections discussed here; see Fig. \ref{ugriSelections}.

%============= ============= ============= ========== 
%=============		  Conclusion		  ===============x%=============== ===================  ===============
%\section{Conclusion}
\section{Conclusion}
We present an efficient emission-line galaxy selection that can provide a sample from which one can measure the BAO feature in the 2-point correlation function at $z>0.6$. With the photometry available today we can plan for a BAO measure to redshift 1 with the BOSS spectrograph. 

A representative set of photometric surveys that might be available for target selection in a near future on more than 2,000 square degrees are :
\begin{itemize}
\item The Kilo Degree Survey (KIDS)\footnote{http://kids.strw.leidenuniv.nl/} aims at observing 1500 square degrees in the \emph{ugri} bands with $3\sigma$ depth of 24.8, 25.4, 25.2, 24.2 using the VST. %  $u=22.0$, $g=22.2$, $r=22.2$, $i=21.3$
\item the South Galactic Cap U-band Sky Survey\footnote{http://batc.bao.ac.cn/Uband/} (SCUSS) aims a $5 \sigma$ limiting magnitude of $23.0$
\item the Dark Energy Survey (DES) aims at observing 5,000 square degrees in \emph{griz} bands with 10 $\sigma$ depth of 24.6, 24.1, 24.3, 23.9. This survey does not include the \emph{u} band \citep{Abbott_2005,2008MNRAS.386.1219B}.
%\item PanSTARRS\footnote{http://pan-starrs.ifa.hawaii.edu/public/} should provide $30000 \; \mathrm{deg^2}$ in \emph{grizy}-bands to 5 $\sigma$ depth of 24
\item the Large Synoptic Survey Telescope (LSST) \citep{Ivezic_2008} plans to observe 20,000 square degrees in \emph{ugrizy} bands with 5 $\sigma$ depth of 26.1, 27.4, 27.5, 26.8, 26.1, 24.9.
\end{itemize}

Using such deeper photometric surveys and improved pipelines, it should be possible to probe BAO to redshift $z=1.2$ in the next 6 years, {\it e.g.} by the \emph{eBOSS} experiment, and to $z=1.7$ in the next 10 years, {\it e.g.} by PFS-SuMIRE%\footnote{http://sumire.ipmu.jp/en/2652} 
 or \emph{BigBOSS} experiment. %\citep{bigBOSS_2011}.% To investigate the redshift range $[1.2,1.7]$, as the tracers are getting fainter, it is necessary to have a 4-m telescope to allocate redshifts in a small exposure of time.
%The AS3/\emph{eBOSS} experiment proposal, which was recently accepted as a follow-up to BOSS, part of the SDSS-III project. If funded it would measure BAO in the redshift range $[0.6,2.2]$ on an area of about $3000 \; \mathrm{deg}^2$ using emission line galaxies for $[0.6,1.2]$ with a selection inspired from this paper's results and quasars as described in \citet{2012MNRAS.tmp.2305S}. Along with BAO, this survey will use redshift space distortions, weak lensing and Ly$\alpha$ quasars to bring new and competitive cosmological constraints.

%\subsection{Future Photometric Surveys}

%============= ============= ============= ========== 
%====================  Aknowledgements ===============
%=============== ===================  ===============

\section*{Acknowledgements}
Johan Comparat especially thanks Carlo Schimd and Olivier Ilbert for insightful discussions about this observational program and its interpretation.

We thank the SDSS-III/BOSS collaboration for granting us this ancillary program.

Funding for SDSS-III has been provided by the Alfred P. Sloan Foundation, the Participating Institutions, the National Science Foundation, and the U.S. Department of Energy Office of Science. The SDSS-III web site is http://www.sdss3.org/.

SDSS-III is managed by the Astrophysical Research Consortium for the Participating Institutions of the SDSS-III Collaboration including the University of Arizona, the Brazilian Participation Group, Brookhaven National Laboratory, University of Cambridge, Carnegie Mellon University, University of Florida, the French Participation Group, the German Participation Group, Harvard University, the Instituto de Astrofisica de Canarias, the Michigan State/Notre Dame/JINA Participation Group, Johns Hopkins University, Lawrence Berkeley National Laboratory, Max Planck Institute for Astrophysics, Max Planck Institute for Extraterrestrial Physics, New Mexico State University, New York University, Ohio State University, Pennsylvania State University, University of Portsmouth, Princeton University, the Spanish Participation Group, University of Tokyo, University of Utah, Vanderbilt University, University of Virginia, University of Washington, and Yale University. 

Based on observations obtained with MegaPrime/MegaCam, a joint project of CFHT and CEA/DAPNIA, at the Canada-France-Hawaii Telescope (CFHT) which is operated by the National Research Council (NRC) of Canada, the Institut National des Science de l'Univers of the Centre National de la Recherche Scientifique (CNRS) of France, and the University of Hawaii. This work is based in part on data products produced at TERAPIX and the Canadian Astronomy Data Centre as part of the Canada-France-Hawaii Telescope Legacy Survey, a collaborative project of NRC and CNRS.

The BOSS French Participation Group is supported by Agence Nationale de la Recherche under grant ANR-08-BLAN-0222.

%----------------- Figure 11. gri selection SDSS CFHT ----------------------------

\bibliographystyle{mn2e}
\bibliography{biblio.bib}
%============= ============= =============  ===========%=====================   Appendixes ====================
%=============== ===================  ================
\appendix 
%======================================================%===============		4.Physics of ELGs		===========
%======================================================

%\section{Table of a subsample of observed galaxies at $z>0.6$}
\section{Table of a subsample of observed galaxies at $z>0.6$}
\label{tble_appendix}
%------------------------- Table B.1: Properties of 30 ELGs -----------------------
\begin{landscape}
\begin{table}
\label{gri_table}
\caption{43 ELGs from the \emph{gri} selection and their observed properties. The error on the redshift determined by the software is not yet fully reliable and thus not displayed. The complete ELG dataset will be included in SDSS Data Release 9 in July 2012.}
\centering
\include{gri_table_2}
\end{table}
\end{landscape}

%------------------------- Figure C.1: ELGs at $z<0.5$ ----------------
\begin{figure*}
\begin{center}
\caption{\emph{ugr}-selected ELGs at $z<0.5$. The spectra are represented in $f_\lambda$ convention. The coloured image on the left is courtesy of CFHT cutout service. The black and white image in the center is the \emph{g}-band. The observed object is the one in the centre of the images.}
\includegraphics[width=180mm]{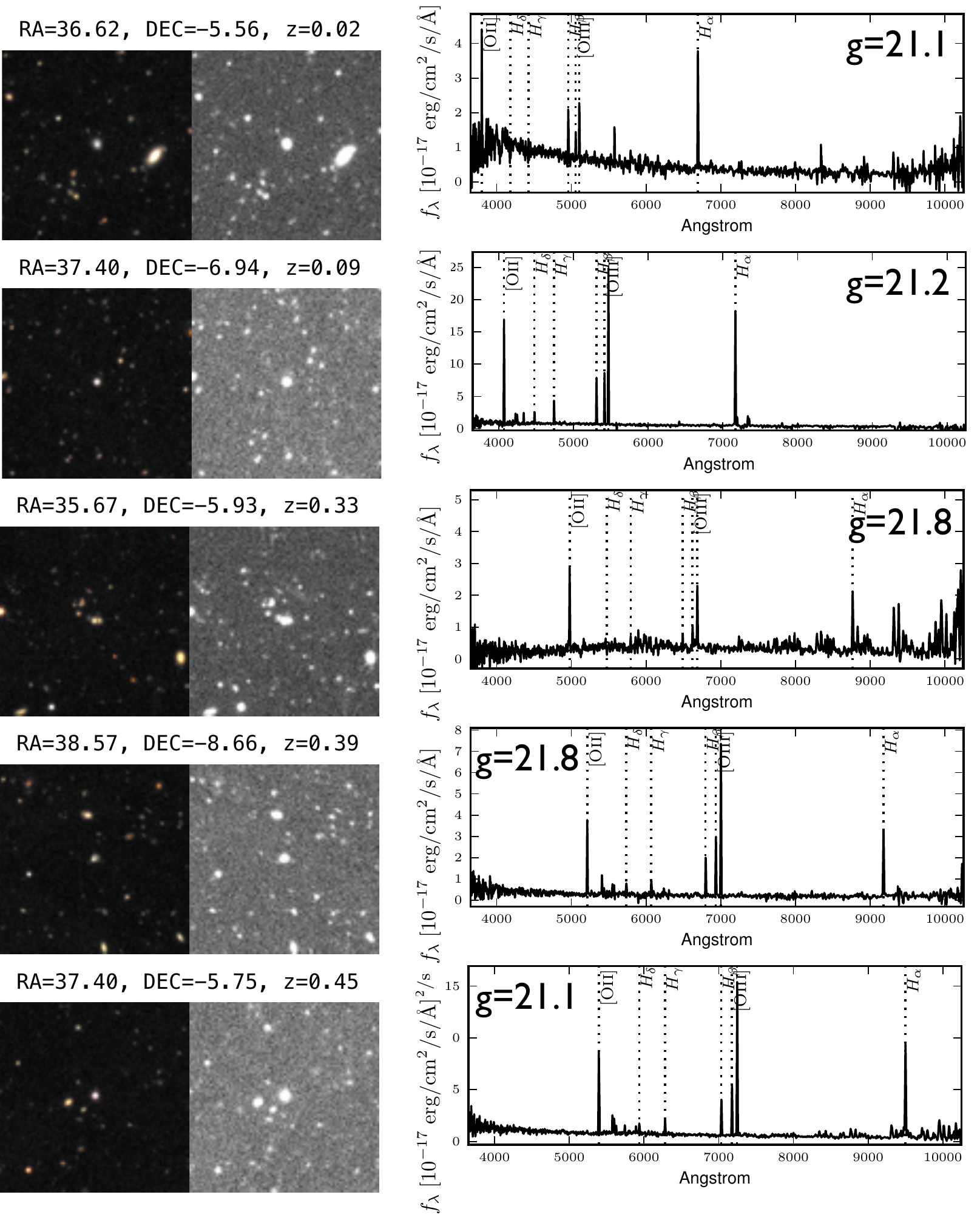}
\label{Em_loz}
\end{center}
\end{figure*}
%------------------------- Figure C.2: ELGs at $0.5<z<1$ ----------------
\begin{figure*}
\begin{center}
\caption{\emph{ugr}-selected ELGs at $0.5<z<1.0$. The spectra are represented in $f_\lambda$ convention. The coloured image on the left is courtesy of CFHT cutout service. The black and white image in the center is the \emph{g}-band. The observed object is the one in the centre of the images.}
\includegraphics[width=180mm]{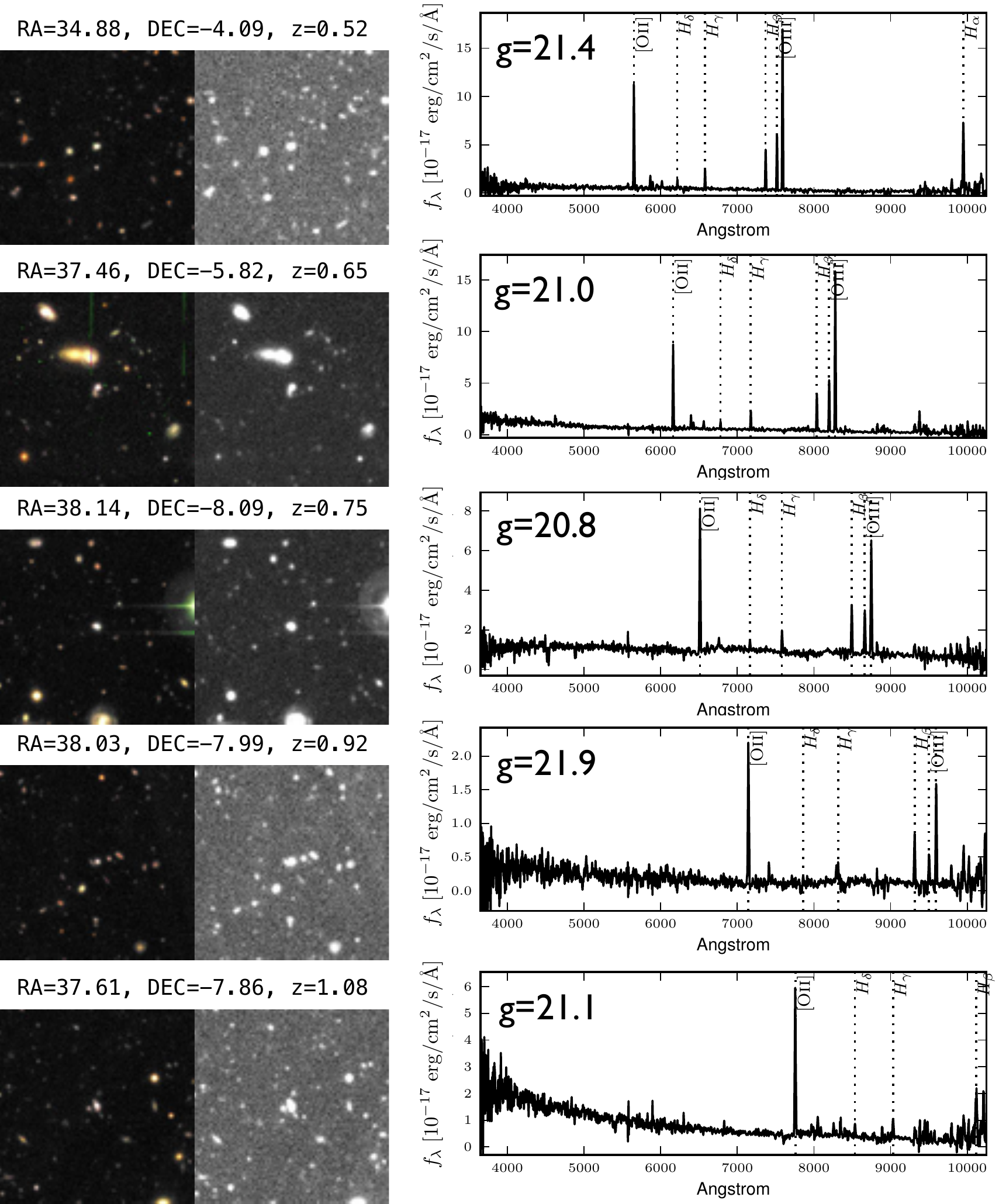}
\label{Em_midz}
\end{center}
\end{figure*}
%------------------------- Figure C.3: ELGs at $1<z<1.6$ ----------------
\begin{figure*}
\begin{center}
\caption{\emph{ugr}-selected ELGs at $1.0<z<1.6$. The spectra are represented in $f_\lambda$ convention. The coloured image on the left is courtesy of CFHT cutout service. The black and white image in the center is the \emph{g}-band. The observed object is the one in the centre of the images.}
\includegraphics[width=180mm]{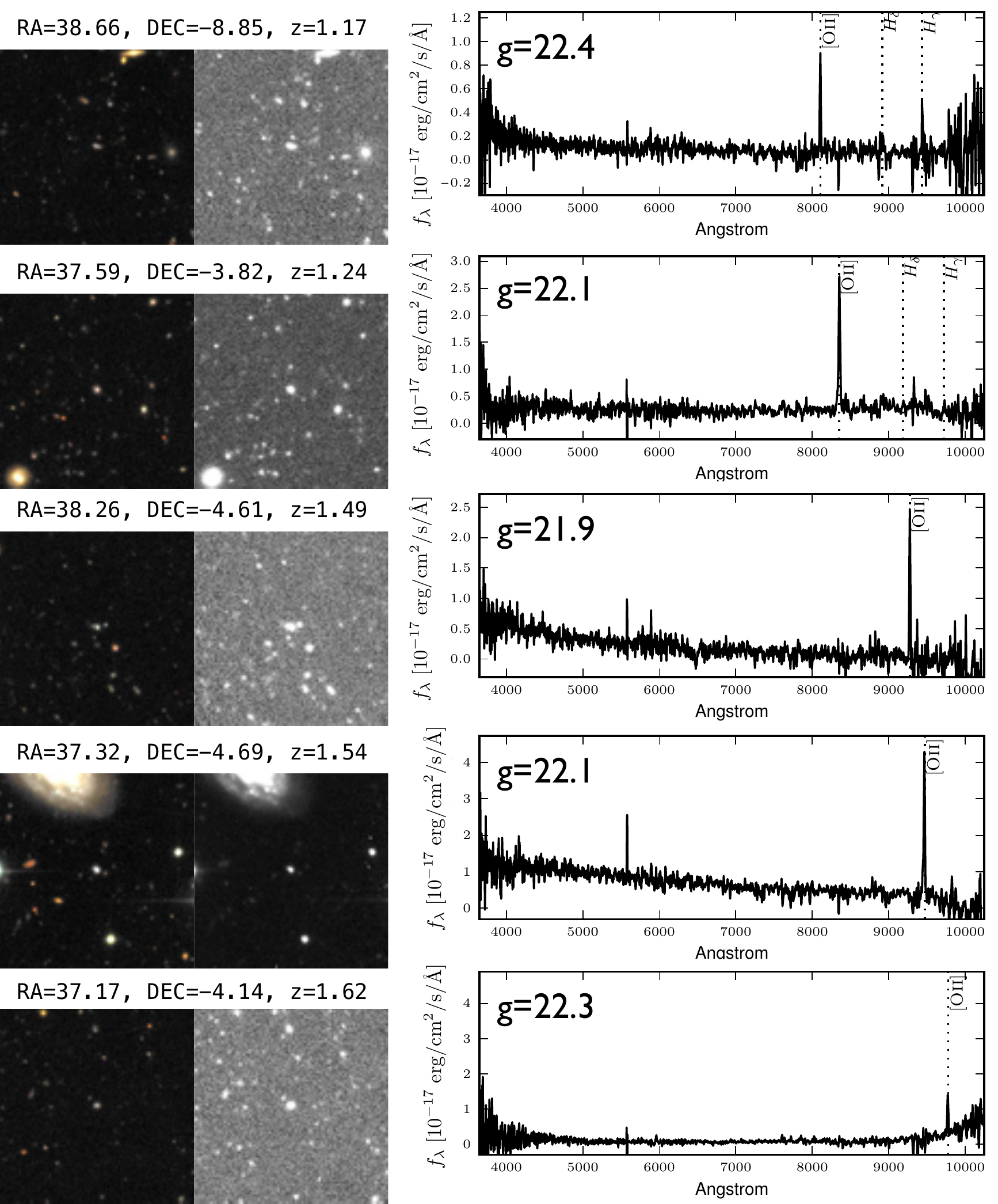}
\label{Em_hiz}
\end{center}
\end{figure*}
%------------------------- Figure C.4: QSOs at $1.0<z<2.3$ ----------------
\begin{figure*}
\begin{center}
\caption{\emph{ugr}-selected quasars at $1.0<z<2.3$. The spectra are represented in $f_\lambda$ convention. The coloured image on the left is courtesy of CFHT cutout service. The black and white image in the center is the \emph{g}-band. The observed object is the one in the centre of the images.}
\includegraphics[width=180mm]{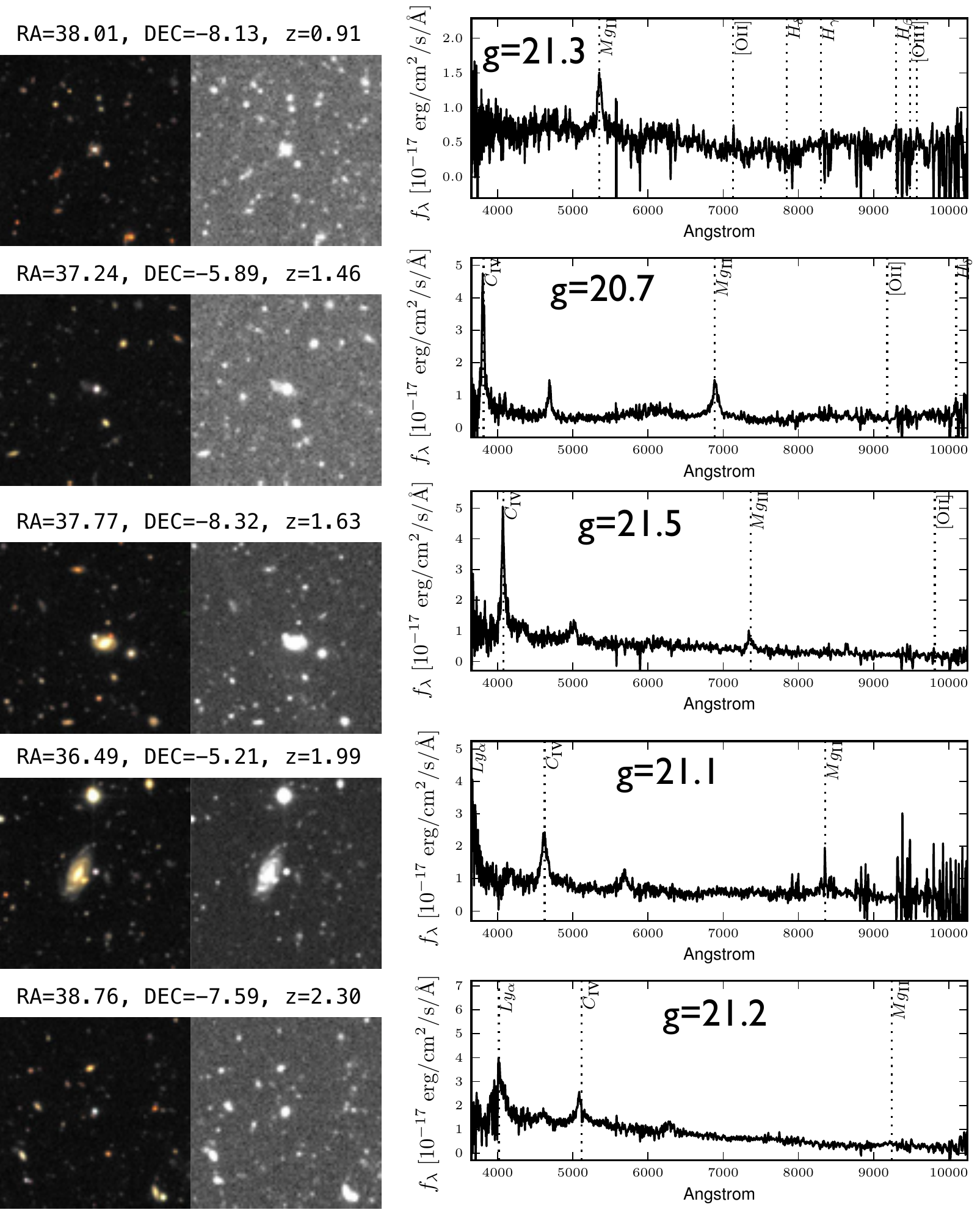}
\label{qsos}
\end{center}
\end{figure*}

\label{lastpage}

\end{document}

%% file: gri_table_2.tex
\begin{tabular}{c c c c c c c c c c c c c c c c c}
%|r |r|r|r| r|r| r| r|r| r| r| r|r| r|r| r| r|}
\hline \hline
  \multicolumn{1}{c}{RA (J2000)} &
  \multicolumn{1}{c}{DEC (J2000)} &
  \multicolumn{1}{c}{redshift} &
  \multicolumn{1}{c}{g} &
  \multicolumn{1}{c}{r} &
  \multicolumn{1}{c}{i} &
  \multicolumn{1}{c}{$r_2$} &
  \multicolumn{1}{c}{$r_2^{limit}$} &
  \multicolumn{1}{c}{$logM_*$} &
  \multicolumn{1}{c}{$EW_{\left[\mathrm{O\textrm{\textsc{ii}}}\right]}^{3727}$} &
  \multicolumn{1}{c}{$flux_{\left[\mathrm{O\textrm{\textsc{ii}}}\right]}^{3727}$} &
  \multicolumn{1}{c}{$EW_{\left[\mathrm{O\textrm{\textsc{iii}}}\right]}^{5007}$} &
  \multicolumn{1}{c}{$flux_{\left[\mathrm{O\textrm{\textsc{iii}}}\right]}^{5007}$} &
  \multicolumn{1}{c}{$D_{4n}$} &
  \multicolumn{1}{c}{$12 \log(OH)$} &
  \multicolumn{1}{c}{$\log (\mathrm{SFR_{H_\beta}})$} &
  \multicolumn{1}{c}{$\log (\mathrm{SFR_{\left[\mathrm{O\textrm{\textsc{ii}}}\right]}})$} \\
\hline
  213.6918335 & 52.7059669 & 0.6055 & 22.79 & 22.13 & 21.28 & 3.46 & 2.4 & 10.5 & -15.0 & 14.15 & -3.98 & 4.57 & 1.12 & 8.77 & 0.35 & 0.68\\
  214.8627014 & 51.775528 & 0.6057 & 22.96 & 21.87 & 20.85 & 3.16 & 3.04 & 10.55 & -4.1 & 6.5 & -3.06 & 5.37 & 1.21 & 9.05 & 0.61 & 0.18\\
  214.234848 & 53.0964775 & 0.6074 & 23.0 & 22.29 & 21.18 & 4.16 & 2.69 & 10.35 & -13.97 & 12.06 & -5.67 & 4.43 & 1.45 & 8.65 & 0.39 & 0.58\\
  213.7070465 & 52.2989693 & 0.6169 & 22.2 & 21.52 & 20.65 & 5.11 & 2.62 & 10.52 & -6.95 & 7.76 & -11.71 & 18.04 & 1.27 & 9.0 & 0.76 & 0.31\\
  216.1340179 & 53.6467667 & 0.6216 & 22.62 & 21.72 & 20.79 & 4.03 & 2.63 & 10.67 & -14.3 & 20.84 & -8.74 & 9.13 & 1.11 & 8.92 & 0.83 & 0.96\\
  213.3887024 & 53.5097351 & 0.6233 & 21.97 & 21.3 & 20.28 & 5.98 & 2.16 & 10.98 & -17.69 & 19.41 & -1.75 & 3.02 & 1.15 & 9.19 & 1.26 & 0.91\\
  215.3585205 & 52.8954582 & 0.6319 & 22.71 & 22.06 & 21.24 & 4.2 & 2.85 & 10.41 & -6.48 & 5.94 & -5.89 & 5.49 & 1.05 & 9.09 & 0.54 & 0.17\\
  213.268158 & 53.4914665 & 0.6325 & 22.79 & 21.73 & 20.71 & 3.06 & 2.16 & 10.83 & -3.63 & 3.33 & -2.96 & 5.85 & 1.21 & 9.07 & 0.39 & -0.21\\
  216.0396729 & 53.8974152 & 0.6453 & 22.17 & 21.39 & 20.52 & 3.75 & 2.63 & 10.72 & -13.94 & 15.13 & -10.69 & 19.87 & 1.25 & 8.86 & 0.85 & 0.79\\
  213.7992401 & 53.8096962 & 0.6486 & 22.78 & 21.81 & 20.78 & 3.96 & 2.69 & 10.62 & -10.3 & 10.82 & -9.41 & 12.07 & 1.26 & 9.03 & 0.8 & 0.58\\
  213.3646545 & 52.2972374 & 0.6516 & 22.55 & 21.97 & 20.8 & 5.33 & 2.4 & 10.65 & -6.27 & 8.9 & -8.67 & 9.83 & 1.05 & 8.89 & 0.48 & 0.46\\
  214.332962 & 52.1850281 & 0.6713 & 22.92 & 22.1 & 21.13 & 3.39 & 3.04 & 10.58 & -5.77 & 6.99 & -11.25 & 19.21 & 1.35 & 8.73 & 0.69 & 0.34\\
  215.8718262 & 52.804966 & 0.6728 & 22.86 & 21.97 & 21.06 & 4.51 & 2.85 & 10.39 & -22.57 & 18.81 & -14.13 & 14.35 & 1.38 & 8.61 & 0.64 & 0.98\\
  216.0170441 & 53.2218399 & 0.6745 & 22.73 & 21.77 & 20.74 & 4.43 & 2.63 & 10.67 & -6.25 & 6.94 & -9.26 & 13.23 & 1.32 & 8.79 & 0.42 & 0.34\\
  213.0988007 & 53.3722191 & 0.7115 & 22.2 & 21.58 & 20.63 & 3.56 & 2.16 & 10.83 & -6.5 & 10.66 & -5.93 & 10.26 & 1.44 & 8.95 & 0.84 & 0.67\\
  213.3162994 & 52.2124023 & 0.7172 & 22.38 & 22.12 & 21.16 & 4.59 & 2.4 & 10.09 & -28.72 & 23.85 & -17.58 & 14.36 & 1.31 & 8.86 & 0.95 & 1.2\\
  216.207016 & 53.3441353 & 0.7214 & 22.68 & 21.65 & 20.75 & 6.29 & 2.63 & 10.91 & -5.67 & 7.19 & -12.6 & 21.03 & 1.33 & 8.93 & 0.92 & 0.43\\
  215.2261353 & 54.4270706 & 0.7232 & 22.51 & 21.75 & 20.84 & 4.74 & 2.77 & 10.91 & -15.3 & 23.72 & -15.91 & 29.68 & 1.33 & 8.98 & 1.22 & 1.2\\
  214.0466309 & 54.2588539 & 0.7332 & 22.38 & 21.85 & 20.88 & 3.56 & 2.94 & 10.34 & -15.25 & 12.34 & -14.08 & 18.21 & 1.05 & 9.1 & 1.2 & 0.8\\
  216.872467 & 53.8220406 & 0.7384 & 22.52 & 21.76 & 20.84 & 4.64 & 2.69 & 11.05 & -8.17 & 8.28 & -6.1 & 11.15 & 1.27 & 9.07 & 1.19 & 0.55\\
  215.8093567 & 52.568676 & 0.7435 & 22.07 & 21.24 & 20.22 & 3.54 & 2.85 & 11.07 & -8.75 & 12.48 & -10.71 & 25.15 & 1.29 & 8.75 & 0.87 & 0.82\\
  214.6755829 & 52.3814735 & 0.7474 & 22.72 & 21.87 & 20.91 & 3.66 & 2.62 & 10.78 & -37.32 & 41.68 & -4.67 & 11.07 & 1.19 & 8.87 & 1.09 & 1.6\\
  214.9938507 & 53.0233078 & 0.7536 & 22.37 & 21.57 & 20.54 & 3.85 & 2.62 & 11.15 & -10.54 & 10.98 & -3.2 & 5.18 & 1.28 & 9.18 & 1.15 & 0.75\\
  216.4916077 & 52.4496536 & 0.7655 & 22.24 & 21.59 & 20.64 & 4.69 & 2.85 & 11.01 & -5.49 & 8.42 & -3.18 & 5.86 & 1.38 & 9.04 & 0.72 & 0.59\\
  215.0812225 & 54.1586342 & 0.7663 & 22.58 & 22.17 & 21.1 & 3.44 & 2.94 & 10.56 & -11.64 & 16.8 & -6.87 & 9.62 & 1.13 & 8.9 & 1.06 & 1.04\\
  213.9476013 & 52.7977905 & 0.7667 & 22.56 & 21.56 & 20.34 & 3.31 & 2.62 & 10.84 & -1.5 & 1.94 & -2.29 & 4.6 & 1.38 & 9.09 & 0.8 & -0.35\\
  214.9521484 & 53.7672997 & 0.7691 & 22.91 & 21.95 & 20.86 & 3.44 & 2.69 & 10.7 & -8.77 & 11.81 & -2.11 & 3.95 & 1.36 & 8.59 & 0.48 & 0.82\\
  215.9055939 & 53.82481 & 0.7727 & 21.84 & 20.94 & 19.96 & 7.69 & 2.63 & 10.75 & -22.57 & 16.87 & -6.3 & 8.81 & 1.18 & 8.91 & 0.91 & 1.05\\
  213.0868225 & 53.8581505 & 0.7814 & 22.47 & 21.93 & 21.09 & 5.33 & 2.16 & 10.65 & -26.32 & 18.35 & -20.79 & 12.91 & 1.22 & 8.76 & 0.92 & 1.12\\
  214.3099976 & 53.5430412 & 0.7861 & 22.19 & 21.75 & 20.93 & 5.76 & 2.69 & 10.54 & -10.87 & 9.3 & -5.97 & 9.67 & 1.1 & 9.1 & 1.14 & 0.69\\
  215.021637 & 53.6823692 & 0.7931 & 22.99 & 22.1 & 21.27 & 4.73 & 2.69 & 10.6 & -16.75 & 16.98 & -8.86 & 11.09 & 1.65 & 8.69 & 0.88 & 1.08\\
  215.9687347 & 51.9373779 & 0.7978 & 22.12 & 21.13 & 20.11 & 6.64 & 2.68 & 10.8 & -16.96 & 17.64 & -7.46 & 13.64 & 1.62 & 9.15 & 1.69 & 1.11\\
  217.0994263 & 53.6485252 & 0.8021 & 22.81 & 22.11 & 20.94 & 3.66 & 2.69 & 11.27 & -12.05 & 12.49 & -4.75 & 12.7 & 1.4 & 9.06 & 1.33 & 0.9\\
  216.3505859 & 52.7362518 & 0.8079 & 22.44 & 21.62 & 20.58 & 4.16 & 2.85 & 11.05 & -6.42 & 8.53 & -7.3 & 14.12 & 1.3 & 9.01 & 0.93 & 0.66\\
  215.0271912 & 54.4110947 & 0.8181 & 21.54 & 20.96 & 20.09 & 7.11 & 2.94 & 11.16 & -5.9 & 11.1 & -8.92 & 22.59 & 1.47 & 8.94 & 1.03 & 0.84\\
  216.8689117 & 53.2321129 & 0.8214 & 22.0 & 21.31 & 20.32 & 4.8 & 2.69 & 10.93 & -11.2 & 10.4 & -6.9 & 7.24 & 1.41 & 8.85 & 0.59 & 0.8\\
  215.1800842 & 54.4046021 & 0.8464 & 22.43 & 21.51 & 20.24 & 4.92 & 2.94 & 11.09 & -7.72 & 8.44 & -13.94 & 26.07 & 1.25 & 8.91 & 1.11 & 0.7\\
  216.5852509 & 52.9398422 & 0.8488 & 22.39 & 21.55 & 20.67 & 3.1 & 2.85 & 11.09 & -22.16 & 27.22 & -5.64 & 12.9 & 1.43 & 9.03 & 1.37 & 1.46\\
  215.3106842 & 53.2784233 & 0.8642 & 22.59 & 22.04 & 20.8 & 5.19 & 2.63 & 10.72 & -7.99 & 10.51 & -5.91 & 11.28 & 1.26 & 8.64 & 0.51 & 0.86\\
  212.5664368 & 53.6004372 & 0.9565 & 22.41 & 22.02 & 21.11 & 3.21 & 2.16 & 10.77 & -19.32 & 30.01 & -25.59 & 54.22 & 1.33 & 8.75 & 1.54 & 1.64\\
  216.5693207 & 52.5144348 & 0.9634 & 22.84 & 22.2 & 21.27 & 4.66 & 2.85 & 10.23 & -33.12 & 52.26 & -58.21 & 109.58 & 1.24 & 8.67 & 1.73 & 2.0\\
  215.3971405 & 54.5257492 & 0.9755 & 22.13 & 21.83 & 20.94 & 3.32 & 2.77 & 10.2 & -28.61 & 41.02 & -27.8 & 58.98 & 1.19 & 8.85 & 1.59 & 1.86\\
  214.368927 & 53.5621452 & 1.191 & 22.7 & 22.17 & 20.85 & 2.27 & 2.69 & 11.75 & -9.52 & 24.57 & 0.0 & 0.0 & 1.55 & -Infinity & -Infinity & 1.72\\
\hline\end{tabular}